\documentclass{IEEEtran}
\usepackage{latexsym}
\usepackage{graphicx}
\usepackage{amsfonts,amssymb,amsmath}
\usepackage{hyperref}
\usepackage{cite}
\usepackage{url}
\usepackage{multirow}
\usepackage{notoccite}
\usepackage{algorithmic}
\usepackage{ulem}
\usepackage{amsmath}
\usepackage{amssymb}
\usepackage{amsfonts}
\usepackage{bm}
\usepackage{balance}
\usepackage{textcomp}
\usepackage{scalerel}
\usepackage{color}
\usepackage{subfigure}
\def\BibTeX{{\rm B\kern-.05em{\sc i\kern-.025em b}\kern-.08em
    T\kern-.1667em\lower.7ex\hbox{E}\kern-.125emX}}
\begin{document}
\title{Artificial-Intelligence-Assisted Multi-Modal Terahertz Sensing and Environment Reconstruction}
\author{Yejian Lyu, Zitong Fang, Zhiqiang Yuan, Henk Wymeersch,~\textit{Fellow, IEEE}, and Chong Han,~\textit{Senior Member, IEEE} 
\thanks{

Y. Lyu, Z. Fang, and C. Han are with the Terahertz Wireless Communications Interdisciplinary Research Center (TWC-IRC) at Shanghai Jiao Tong University, Shanghai 200240, China (e-mail: \{yejian.lyu; zitong.fang; chong.han\}@sjtu.edu.cn);

Z. Yuan and H. Wymeersch are with the Communication Systems Group at Chalmers University of Technology, Gothenburg, Sweden (e-mail: {yuanzhiq; henk}@chalmers.se). 

}}
\maketitle
\begin{abstract}
Multi-modal sensing is an important enabler for future environment-aware wireless systems, since a single sensing modality is generally insufficient to provide accurate metric geometry, material awareness, and semantic interpretability in complex environments. This paper presents a measurement-based multi-modal terahertz (THz) sensing and vision framework for indoor environment reconstruction. A three-dimensional monostatic THz channel sounding system operating at 290–310 GHz is integrated with an omnidirectional fisheye camera to acquire radio-frequency and visual observations from a common sensing viewpoint. From the measured THz data, a signal processing pipeline extracts multipath components and infers geometry- and material-consistent structural primitives through trajectory-tracking-assisted parameter estimation, graph-based structure discovery, planar reconstruction, and reflection-loss analysis. In parallel, artificial intelligence (AI)-based visual perception modules extract object-level semantic masks and depth priors from panoramic images. To associate these heterogeneous representations, an agentic-AI-based task-driven THz-agent module is developed to select appropriate integration tools according to the attributes of the modality-specific outputs. Through angular alignment and consistency analysis, THz-derived metric geometry and material information are associated with vision-derived semantic regions and depth priors, enabling geometry-consistent and semantically interpretable environment reconstruction directly from measurements. Experimental validation in the indoor L-shaped hallway demonstrates that the proposed framework reconstructs dominant structural elements with centimeter-level accuracy while identifying semantic categories and material attributes of representative indoor objects. These results show the potential of THz–vision integration for environment-aware sensing, wireless digital twins, and future ISAC systems.
\end{abstract}


\section{Introduction}
\label{sec:introduction}
The global rollout of fifth-generation (5G) networks has unlocked unprecedented connectivity and enabled a wide range of data-intensive applications. However, the relentless pursuit of smarter, more immersive, and context-aware services, such as autonomous mobility, extended reality, and digital twins, has made it clear that 5G will not suffice in meeting the performance and perceptual needs of the next decade. 
In this context, sixth-generation (6G) is envisioned not merely as a faster wireless network but as an intelligent and perceptual infrastructure in which integrated sensing and communication (ISAC) enables communication nodes to reuse the same radio hardware and spectrum for both data transmission and environmental sensing, thereby supporting environment-aware wireless operation, wireless digital twins, embodied intelligent agents, and task-driven local sensing~\cite{6G_isac1,6G_isac2,isac_ris,11146803,11364301, cheng2026apeg}. These applications require not only channel state information or target-level detection results, but also a structured understanding of the surrounding physical environment. For indoor wireless systems, the locations, orientations, and material properties of walls, ceilings, doors, corridors, and major reflecting objects directly affect multipath propagation, blockage conditions, beam selection, and site-specific channel characteristics. Therefore, geometry-consistent environment reconstruction is of practical importance for environment-aware channel modeling, digital-twin construction, beam management, network planning, and adaptive sensing-communication optimization.

Monostatic sensing is particularly relevant for these applications since it uses a co-located transmitter and receiver, leading to a compact and deployment-friendly sensing architecture~\cite{liu2024shared,yejian_thz_mag}. Compared with bistatic or multistatic configurations, monostatic sensing does not require spatially distributed and tightly synchronized transceivers, making it attractive for practical ISAC nodes, mobile sensing platforms, and local environmental mapping systems. At THz frequencies, the large available bandwidth and directional propagation characteristics enable high-resolution range- and angle-domain observations of backscattered paths, which can be exploited to infer dominant environmental structures such as planar surfaces, corners, and major reflecting objects. However, interpreting monostatic THz measurements remains challenging because the surrounding geometry and material properties must be inferred from reflected and multi-bounce propagation paths under realistic propagation conditions.

Despite extensive progress in millimeter wave/THz communications and ISAC-enabled sensing, radar, and target detection, measurement-based ISAC sensing for geometry-consistent environment reconstruction remains comparatively underexplored, particularly for monostatic sensing systems~\cite{mmwave_sensing1,mmwave_sensing3,THz_sensing1,THz_sensing2,Guangzheng2023A}. Existing studies often emphasize target-level sensing or communication-oriented channel characterization, whereas reconstructing structured environmental geometry and material attributes from realistic reflected and multi-bounce THz propagation remains insufficiently addressed. Although monostatic radar channel and echo modeling has been extensively studied in the traditional radar community, such as in automotive and military radar, most ISAC-oriented studies emphasize waveform design, sensing accuracy, and resource allocation, and often rely on simplified or communication-centric channel models that are insufficient to capture realistic sensing propagation effects~\cite{mmwave_sensing1,mmwave_sensing3}. To address this limitation, sensing-aware ISAC channel models have been investigated in various scenarios, including localization-assisted three-dimensional (3D) non-stationary models for dynamic sensing~\cite{yang2023localization}. Beyond scenario-specific studies, several works further extend communication-oriented channel modeling frameworks to incorporate sensing-relevant features, such as additional reflection clusters, radar cross-section modeling, spatial consistency, and shared multipath evolution, particularly for bistatic or network-centric sensing configurations~\cite{zhao2023bistatic,zhang2023shared,liu2024shared,liu2024extend,yang2023novel,yang2024standardization}. Despite these advances, most existing models remain statistically oriented and are predominantly developed for bistatic sensing, offering limited physical interpretability at the level of individual propagation paths. Although recent studies have demonstrated geometry- and material-aware environment mapping using monostatic THz sensing~\cite{lyu2025hybrid,fang2025environment}, multipath components (MPCs) are still largely treated as unlabeled physical entities. Consequently, a key open problem remains the lack of a measurement-based framework that can systematically associate monostatic THz multipath propagation with geometry-consistent and semantically meaningful environmental structures. In this work, ‘semantically meaningful environmental structures’ specifically refer to reconstructed physical components that have both metric geometric attributes and category-level object/structure identities. The metric geometric attributes include position, orientation, and spatial extent, while the category-level identities indicate what each reconstructed component corresponds to in the environment, such as a wall, floor, ceiling, door, corridor boundary, or dominant indoor object. Therefore, the term `semantic' in this work denotes category-level object/structure identification attached to reconstructed geometry, rather than unlabeled geometric reconstruction alone.

Relying solely on a single sensing modality imposes fundamental limitations in perceptual completeness. Visible-light cameras provide rich semantic information, such as object class, material type, and surface details; however, they are inherently constrained by direct line-of-sight (LoS) visibility and their performance degrades significantly under poor illumination or adverse environmental conditions~\cite{guan2020through}. Consequently, vision-only systems struggle to deliver robust and consistent scene understanding in complex indoor environments. In contrast, RF-based sensing techniques, including radar and THz echoes, are well suited for extracting geometric attributes such as distance, angular location, and reflective characteristics by exploiting multipath propagation and material-dependent interactions~\cite{yu2024mobirfpose}. These complementary capabilities of vision and RF sensing have motivated growing interest in multi-modal sensing frameworks, where heterogeneous sensor data are jointly utilized to enhance robustness and perceptual richness~\cite{manjur2025multimodal}. Recent studies have demonstrated that multi-modal integration can significantly improve beam prediction, blockage awareness, and environment mapping in communication-centric sensing systems~\cite{alkhateeb2023deepsense,charan2022vision,mollah2025multimodality,mashhadi2021federated,cheng2023intelligent,zhu2025raytracing,cheng2025synthsom}. Public datasets and frameworks, such as DeepSense 6G~\cite{alkhateeb2023deepsense} and SynthSoM~\cite{cheng2025synthsom}, have further facilitated artificial intelligence (AI)-driven learning from synchronized radio-frequency and visual observations. However, most existing multi-modal systems primarily target communication performance or data-driven optimization, while paying limited attention to the physical interpretability of RF sensing propagation and the semantic grounding of multipath measurements, particularly at high frequencies. This leaves an open challenge in developing tightly integrated multi-modal sensing frameworks that jointly connect RF propagation mechanisms with semantically meaningful environmental structures.

To address these challenges, this paper presents and experimentally validates a multi-modal THz sensing and vision framework for indoor environment reconstruction. The proposed system integrates high-resolution monostatic THz channel sounding with omnidirectional fisheye imaging, enabling the joint acquisition of 3D delay-azimuth-elevation THz measurements and panoramic visual observations from a common sensing viewpoint. This co-located sensing configuration establishes a unified angular-domain representation, which provides a natural interface for geometry-consistent cross-modal association. From the THz measurements, a model-driven processing pipeline extracts spatially consistent multipath components and infers geometry- and material-consistent structural primitives through 3D trajectory-tracking-based parameter estimation, graph-based structure discovery, planar reconstruction, and reflection-loss analysis. In parallel, AI-based vision modules extract object-level semantic categories, angular support regions, and depth priors from panoramic images. Based on these modality-specific outputs, a task-driven cross-modal integration framework associates THz-derived metric geometry and material information with vision-derived semantic regions through angular alignment and consistency analysis.

The key contributions of this paper are summarized as follows:
\begin{itemize}
\item A multi-modal THz-vision sensing system is proposed, which tightly integrates a high-resolution monostatic 0.3~THz channel sounder with an omnidirectional fisheye camera for joint radio-visual sensing in an indoor L-shaped hallway scenario. A comprehensive multi-modal measurement campaign is conducted at $15$ sensing locations over a $20$~GHz bandwidth, where 3D delay-azimuth-elevation THz monostatic measurements and panoramic visible-light images are acquired in a geometry-consistent manner from the same physical viewpoint.
\item A THz sensing and parameter extraction pipeline based on model-based signal processing methods is developed to infer geometry- and material-consistent structural primitives from measured multipath components. In parallel, an AI-based visual perception module is employed to process panoramic visible-light images and extract category-level object/structure information. Besides, an AI-based depth-estimation module is introduced to estimate depth maps from the panoramic images, which provide auxiliary geometric priors for subsequent multi-modal integration.
\item A task-driven THz-agent framework is proposed to jointly integrate THz-derived geometric and material parameters with vision-derived semantic labels in an angle-conditioned manner. By exploiting the complementary geometry, material, structural, and semantic information obtained from multi-modal measurements, the proposed framework enables geometry-consistent and semantically interpretable indoor environment reconstruction.
\end{itemize}

The remainder of this paper is structured as follows. Section~\ref{sec:sounder} outlines the monostatic-sensing-based measurement setup and scenario. The AI-assisted signal processing methods for each modality are described in Section~\ref{sec:estimator}. The cross-model integration and environment reconstruction are analyzed in Section~\ref{sec:material}. Finally, concluding remarks are presented in Section~\ref{sec:conclusion}.

\section{Measurement Campaign}\label{sec:sounder}
\begin{figure}
\centering
	\subfigure []{\includegraphics[width=0.42\columnwidth]{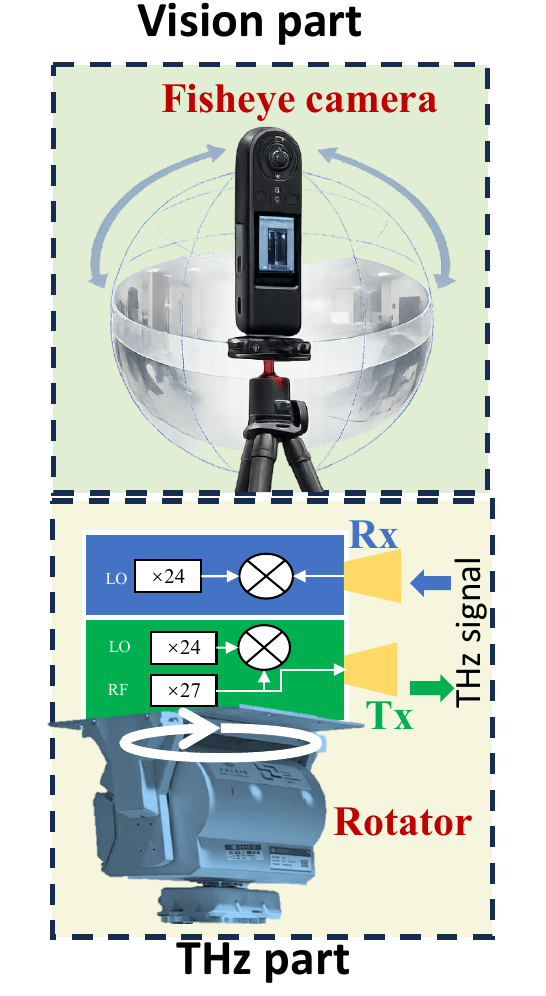}} 
	\subfigure []{\includegraphics[width=0.52\columnwidth]{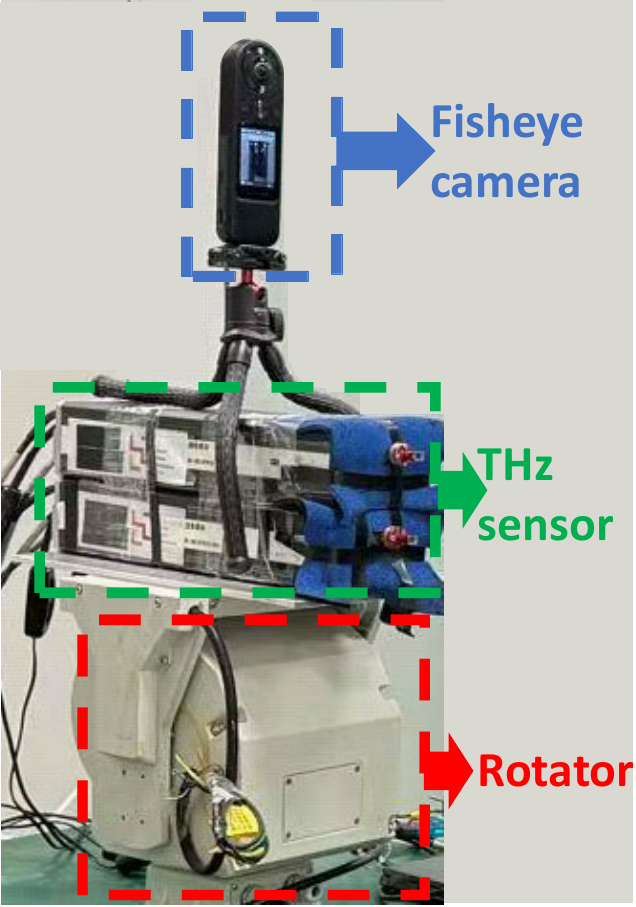}}
\caption {The multi-modal sensing system. (a) Schematic diagram. (b) A picture of the system.}\label{fig:system_pic}
\end{figure}

In this work, the proposed multi-modal sensing system consists of two complementary components, namely a THz monostatic sensing subsystem and a panoramic visible-light imaging subsystem, as illustrated in Fig.~\ref{fig:system_pic}. The THz sensing subsystem is based on a vector network analyzer (VNA)-based channel sounder~\cite{yejian_mag} operating in the frequency range from $290$ to $310$~GHz. By exploiting an ultra-wide bandwidth of $20$~GHz, the system achieves a fine delay resolution of approximately $1.5$~cm in the delay domain. A total of $8001$ frequency points are sampled, which supports a maximum unambiguous sensing range of $120$~m. To ensure high dynamic range and measurement accuracy, the intermediate-frequency (IF) bandwidth is set to $1$~kHz. Identical high-gain horn antennas are employed at both the transmitting and receiving ports, featuring a gain of $25.5$~dBi and narrow half-power beamwidths (HPBWs) of $8^{\circ}$ in azimuth and $6^{\circ}$ in elevation, thereby enabling high angular resolution. The transmitter (Tx) and receiver (Rx) are co-located, forming a monostatic sensing configuration.

The visible-light sensing subsystem consists of a fixed-position, omnidirectional fisheye camera (QooCam~8K) that captures high-resolution panoramic images of the measurement environment. With a full $360^{\circ}$ horizontal field of view and near-complete vertical coverage, the fisheye camera observes the entire surrounding scene from a single static viewpoint, without requiring mechanical rotation, scanning, or multiple cameras. This enables comprehensive visual observation of scene geometry and object distribution within the sensing area. The captured panoramic images serve as a complementary modality to the THz sensing data, supporting environment annotation, geometric reference extraction, and visual verification of dominant scatterers and reflecting surfaces. In particular, the high angular resolution of the fisheye imagery is well suited for associating visually observable structures—such as walls, doors, fire hydrant boxes, and other indoor objects—with corresponding MPCs identified in the THz monostatic measurements.

As shown in Fig.~\ref{fig:system_pic}, the THz sensor and the fisheye camera are co-mounted on a precision-controlled turntable to ensure accurate and repeatable angular coordinate alignment between the radio-frequency and visual sensing modalities. Note that only the THz antenna requires mechanical rotation during the measurement process, while the fisheye camera remains static and captures a single panoramic image at each sensing location. The monostatic THz sensor is deployed at two heights, namely $2.0$~m and $2.2$~m, to capture height-dependent propagation characteristics. At each sensing location, the measurement procedure begins with the acquisition of a panoramic visible-light image, followed by THz sensing measurements. For the THz subsystem, a directional scanning sounding (DSS) scheme is adopted, in which the antenna is rotated over an azimuth range of $[0^{\circ}, 355^{\circ}]$ with a step size of $5^{\circ}$ and an elevation range of $[-20^{\circ}, 20^{\circ}]$ with the same angular resolution, enabling dense sampling of the 3D angular domain. A summary of the key system and measurement parameters is provided in Table~\ref{tab:configuration}.

\begin{table}
\centering
\caption{The multi-modal sensing measurement configuration.}\label{tab:configuration}
\begin{tabular}{c|c}
\hline
Parameter   &  Value\\
\hline
\hline
Frequency range & $290$-$310$~GHz\\
\hline
Frequency bandwidth      & $20$~GHz\\
\hline
Frequency point      & $8001$\\
\hline
Transmission power &  $10$~dBm\\
\hline
Noise floor &  $-170$~dB\\
\hline
Maximum detectable distance &  $120$~m\\
\hline
IF bandwidth            & $1$~kHz\\
\hline
Tx \& Rx antenna type   & Horn\\
\hline
Antenna gain   & $25.5$~dBi\\
\hline
Azimuthal HPBW           & $8^{\circ}$\\
\hline
Elevational HPBW           & $6^{\circ}$\\
\hline
Rotation step     & $5^{\circ}$\\
\hline
Azimuthal rotation range  & $[0^{\circ}:5^{\circ}:355^{\circ}]$\\
\hline
Elevational rotation range  & $[-20^{\circ}:5^{\circ}:20^{\circ}]$\\
\hline
Sensor height     & $2.0$~m\\
\hline
Camera type     & QooCam 8k\\
\hline
Camera height     & $2.2$~m\\
\hline
\end{tabular}
\end{table}

\begin{figure}
    \centering
    \includegraphics[width=1\columnwidth]{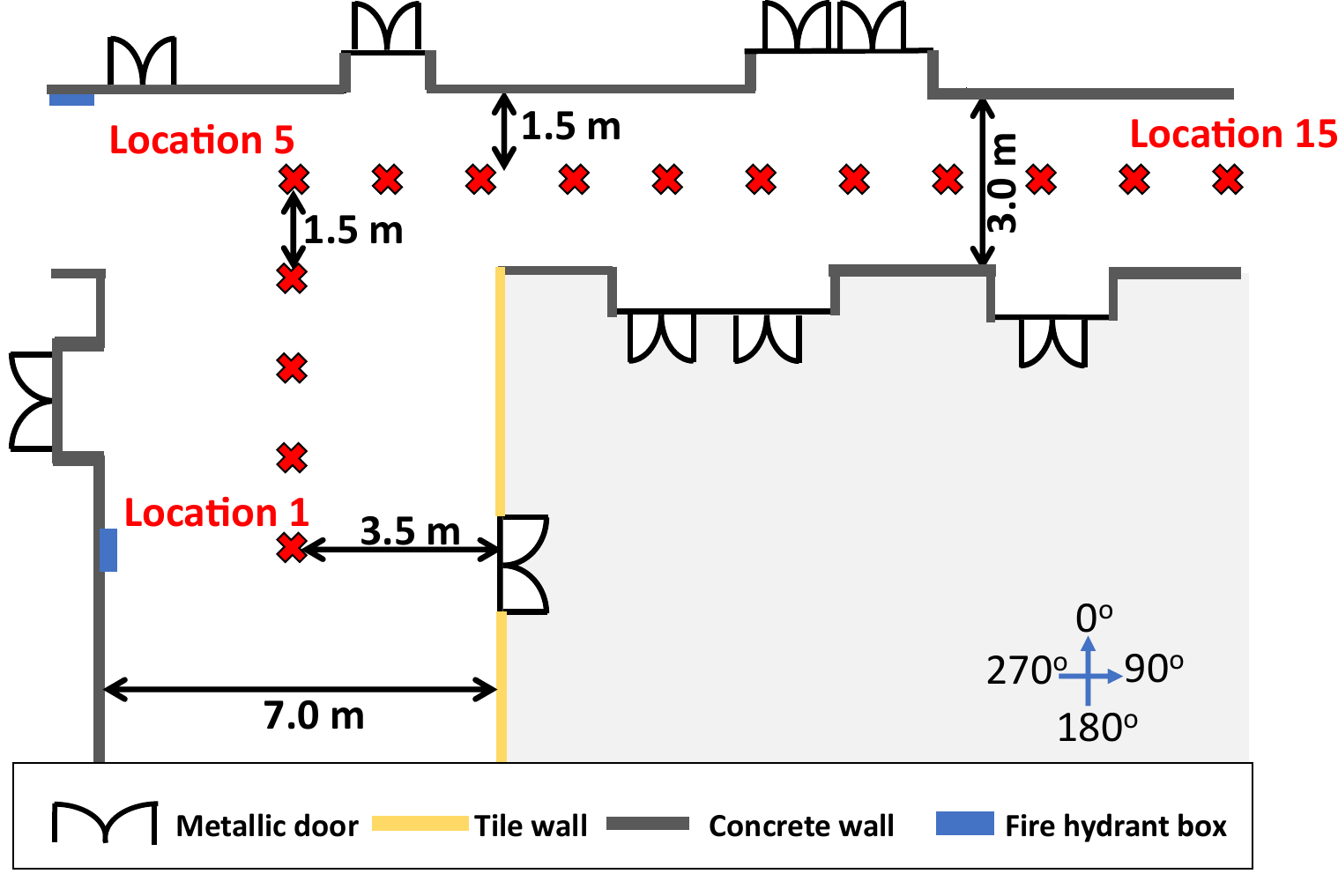}
\caption {The schematic diagram of the scenario and the deployment of the multi-modal transceiver.}\label{fig:scenario_pic} 
\end{figure}

The multi-modal measurement campaign is conducted in an indoor L-shaped hallway scenario, as depicted in Fig.~\ref{fig:scenario_pic}. The scenario comprises a diverse set of common indoor materials, including concrete walls, ceramic tile surfaces, 2 fire hydrant boxes, and 9 metallic doors, thereby offering rich and heterogeneous reflection characteristics for THz propagation analysis. Along the L-shaped route, a total of $15$ sensing locations are selected with a uniform spacing of $1.5$~m between adjacent positions. At each location, the DSS scheme yields $72$ azimuth directions and $9$ elevation angles, resulting in a total of $15 \times 72 \times 9 = 9720$ channel impulse response measurements. The measurement duration at each sensing location is approximately $4$~hours, leading to an overall campaign time of about $60$~hours. Note that this acquisition time is mainly due to the dense $3$D mechanical scanning required by the VNA-based THz sounder with highly directive antennas. Therefore, the present prototype is intended for offline high-resolution reconstruction in static or quasi-static environments, such as site surveying, indoor digital-twin generation, infrastructure inspection, and after-hours office or factory mapping, rather than real-time dynamic sensing. For future practical deployment, the acquisition time can be reduced by using faster time-domain or correlation-based THz sounders, electronically steerable arrays, sparse/adaptive angular sampling, and camera-guided region-of-interest scanning.

In parallel with the THz measurements, a single panoramic visible-light image is captured at each sensing location using the QooCam~8K fisheye camera operating in a static capture mode. The camera provides omnidirectional imagery with a maximum resolution of $7680 \times 3840$ pixels, corresponding to an effective horizontal angular resolution of approximately $0.047^{\circ}$ per pixel. As a result, the visual dataset consists of $15$ high-resolution panoramic images, yielding a total data volume on the order of several hundred megabytes, depending on the compression format. Compared with the THz channel measurements, the visual data volume is relatively modest, while providing rich semantic and geometric information that supports subsequent multi-modal analysis and integration.

\section{AI-Assisted Processing for Each Modality}\label{sec:estimator}
This section presents the AI-based processing methods developed for the measured multi-modal sensing data. It first introduces the signal model and parameter extraction procedure for monostatic THz sensing, followed by an unsupervised AI-based processing pipeline for extracting geometry-consistent multipath structures. Subsequently, the vision-based AI method for panoramic semantic segmentation is described, which provides object-level semantic information to support multi-modal environment reconstruction.

\subsection{Signal Processing for THz}\label{subsec:trajectory_tracking}
\subsubsection{Signal Model}
To extend the signal model from the delay-azimuth domain to the delay-azimuth-elevation domain, both the directional channel frequency response (CFR) and the associated rotation manifold are generalized to explicitly account for elevation. At the $m^{\text{th}}$ measurement location, the directional CFR observed at carrier frequency $f$ with azimuth and elevation rotation angles $(\theta,\vartheta)$ can be expressed as
\begin{align}
	H_{m}(f,\theta,\vartheta)
	= \sum_{\ell=1}^{L_m} \alpha_{\ell}^{(m,\theta,\vartheta)}
	\exp\!\left(-j2\pi f \tau_{\ell}^{(m)}\right)
	\, a_{{\rm TRx},\ell}(f,\theta,\vartheta),
	\label{equ:signal_3d}
\end{align}
where $\alpha_{\ell}^{(m)}$ and $\tau_{\ell}^{(m)}$ denote the complex amplitude and propagation delay of the $\ell^{\text{th}}$ MPCs, respectively, and $L_m$ represents the total number of MPCs at the $m^{\text{th}}$ location. Consistent with the mimic monostatic sensing configuration considered in this work, the azimuth and elevation angles of departure and arrival are assumed to be identical for each MPC.

The term $a_{{\rm TRx},\ell}(f,\theta,\vartheta)$ denotes the 3D rotation manifold coefficient, which incorporates both the phase variation induced by the DSS mechanism and the complex 3D radiation pattern of the transceiver. It can be written as
\begin{align}
	a_{{\rm TRx},\ell}(f,\theta,\vartheta)
	& = \Big[\exp\!\big(j2\pi f\, \Delta_{\ell}(\theta,\vartheta)/c\big)\Big]^2 \nonumber\\
	& G_{\rm TRx}\!\big(f,\phi_{\ell}^{(m)}-\theta,\psi_{\ell}^{(m)}-\vartheta\big),
	\label{equ:manifold_3d}
\end{align}
where $c$, $\phi_{\ell}^{(m)}$, and $\psi_{\ell}^{(m)}$ denote the speed of light, azimuth angle, and elevation angle of the $\ell^{\text{th}}$ MPC, respectively. And $G_{\rm TRx}(\cdot)$ is the complex 3D radiation pattern of the mimic monostatic sensor, evaluated at the relative azimuth and elevation offsets between the physical MPC direction and the sensor pointing direction. The DSS-induced excess path length $\Delta_{\ell}(\theta,\vartheta)$ is determined by the projection of the lever-arm distance $r$, defined as the distance between the antenna phase center and the rotation center, onto the MPC direction in 3D space, and can be expressed as
\begin{align}
	\Delta_{\ell}(\theta,\vartheta)
	= r \cos\!\big(\phi_{\ell}^{(m)}-\theta\big)
	\cos\!\big(\psi_{\ell}^{(m)}-\vartheta\big).
\end{align}

The squared exponential term in~\eqref{equ:manifold_3d} accounts for the two-way phase accumulation inherent to monostatic sensing. 

\begin{figure}
	\centering
	\subfigure []{\includegraphics[width=1\columnwidth]{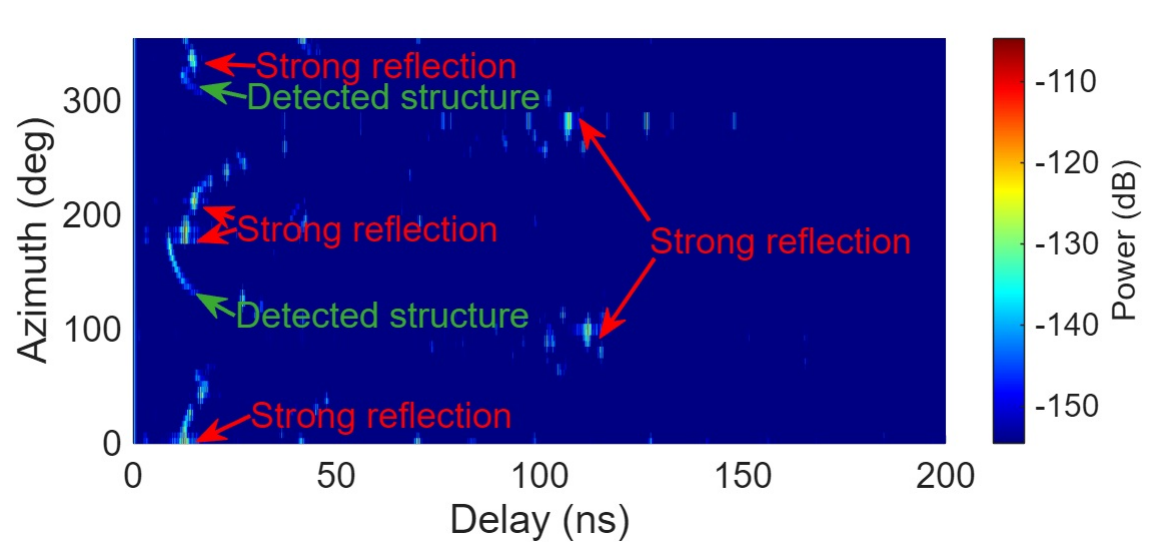}} 
	\subfigure []{\includegraphics[width=1\columnwidth]{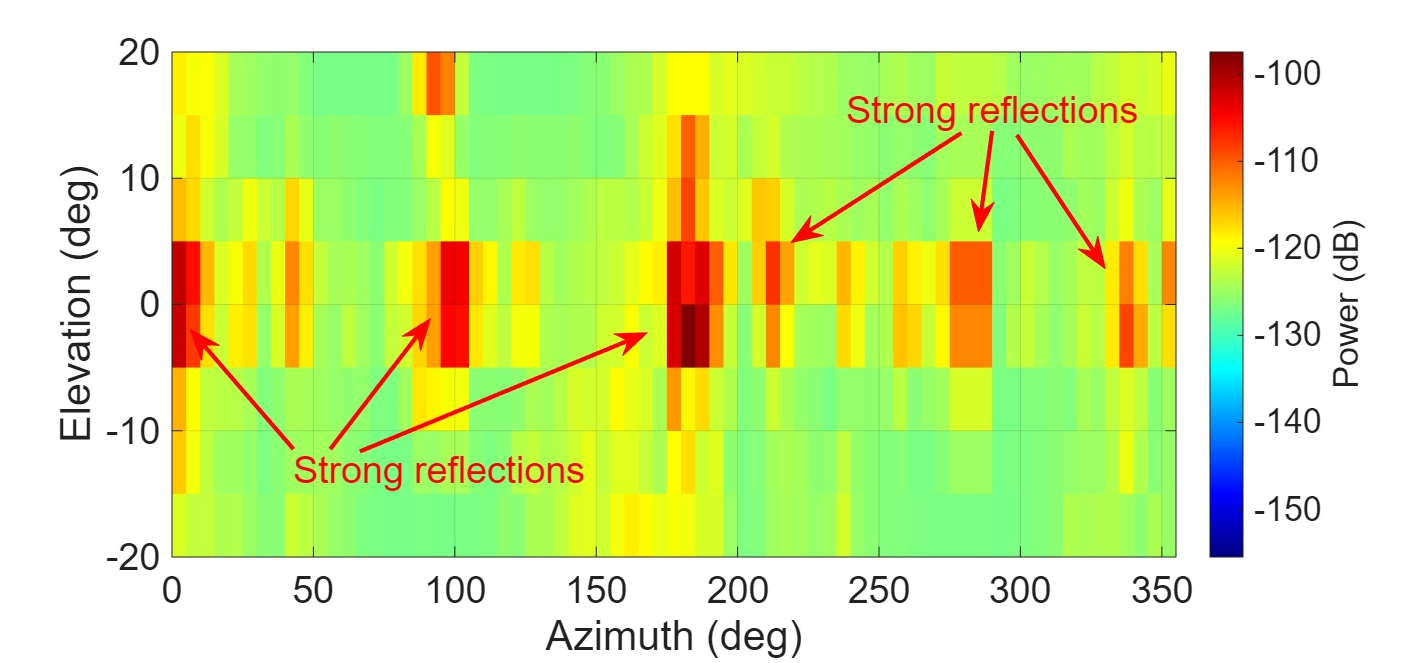}}
	\caption {The example of the THz sensing results in the Location~$11$. (a) PADP result at the elevation of $0^{\circ}$. (b) PAS result.}\label{fig:power_spectra}
\end{figure}

\subsubsection{Data Processing}
During the measurements, directional channel frequency responses (CFRs) $H_m(f,\phi,\varphi)$ are acquired at each monostatic sensing location. By applying an inverse discrete Fourier transform (IDFT) along the frequency dimension, the corresponding directional channel impulse responses (CIRs) $h_m(\tau,\phi,\varphi)$ are obtained, yielding a delay-resolved representation of the backscattered THz signals. Based on the directional CIRs, the power-angle-delay profiles (PADPs) and power-angle spectra (PAS) are constructed by computing the squared magnitude $|h_m(\tau,\phi,\varphi)|^2$. The PADP characterizes the joint distribution of received echo power over propagation delay and angular direction, while the PAS provides a compact angular-domain representation by integrating power contributions across the delay dimension.

Owing to the ultra-wide bandwidth and highly directive antennas, THz monostatic sensing provides high ranging accuracy and fine angular resolution, enabling the detection of dominant MPCs associated with major reflecting structures. As illustrated by the representative PADP and PAS results at Location~11 in Fig.~\ref{fig:power_spectra}~(a) and Fig.~\ref{fig:power_spectra}~(b), strong reflections appear as localized high-power components at distinct delay-angle combinations, while extended reflecting regions are manifested as continuous or distributed angular-domain responses. These results are used as physical delay--angle observations for subsequent MPC extraction and structural reconstruction, rather than as direct semantic labels of specific objects.

However, despite its strong geometric and material sensitivity, THz monostatic sensing alone cannot unambiguously determine object class or semantic meaning. Different objects with similar orientations, surface properties, or material compositions may produce comparable reflection signatures, leading to inherent ambiguities in object-level interpretation. In addition, while THz sensing can provide partial observability of visually occluded structures through electromagnetic penetration of thin or low-loss materials—such as glass panels, paper sheets, or curtains—it does not directly encode object boundaries, shape continuity, or semantic identity. As a result, the physical information obtained from THz sensing, although accurate in distance and material interaction, remains insufficient for reliable semantic understanding of the environment. These limitations motivate the incorporation of complementary semantic information from visible-light perception.
\begin{figure}
	\centering
	\subfigure []{\includegraphics[width=0.9\columnwidth]{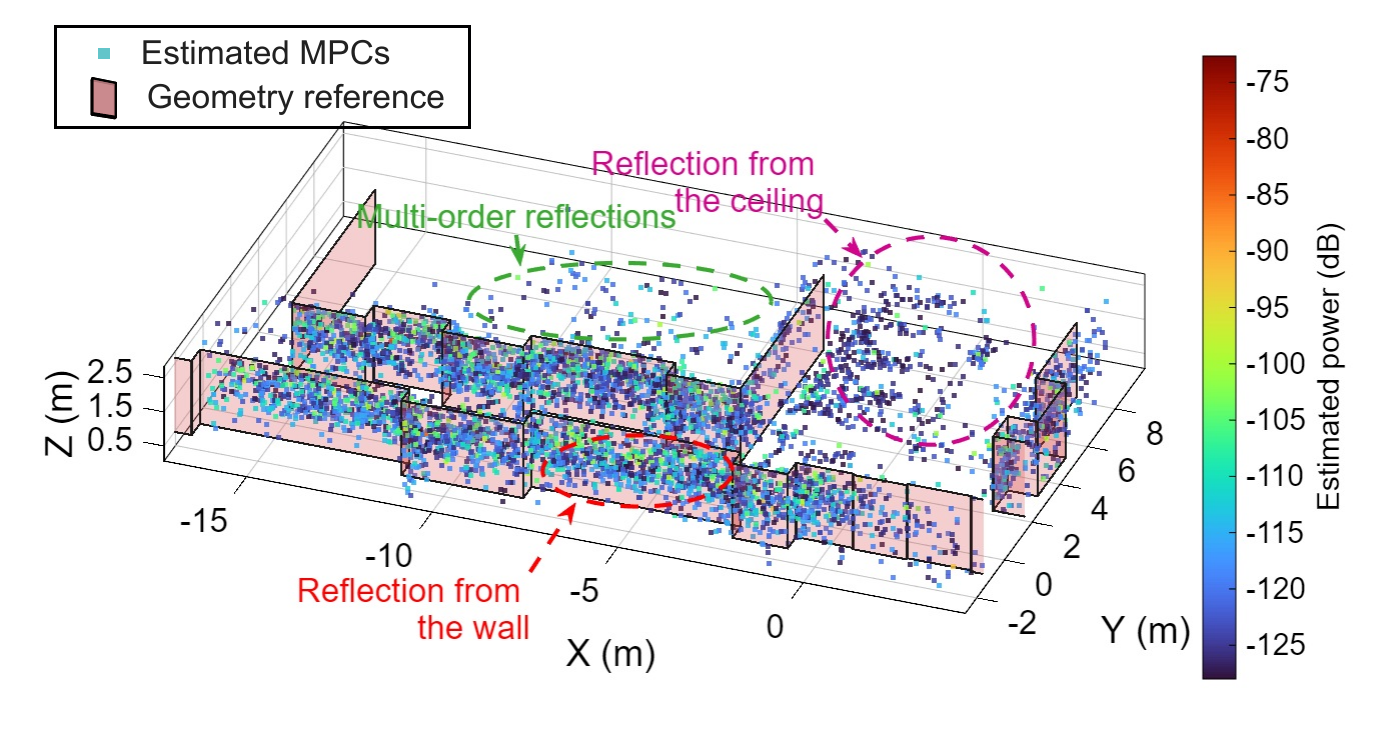}} 
	\subfigure []{\includegraphics[width=0.9\columnwidth]{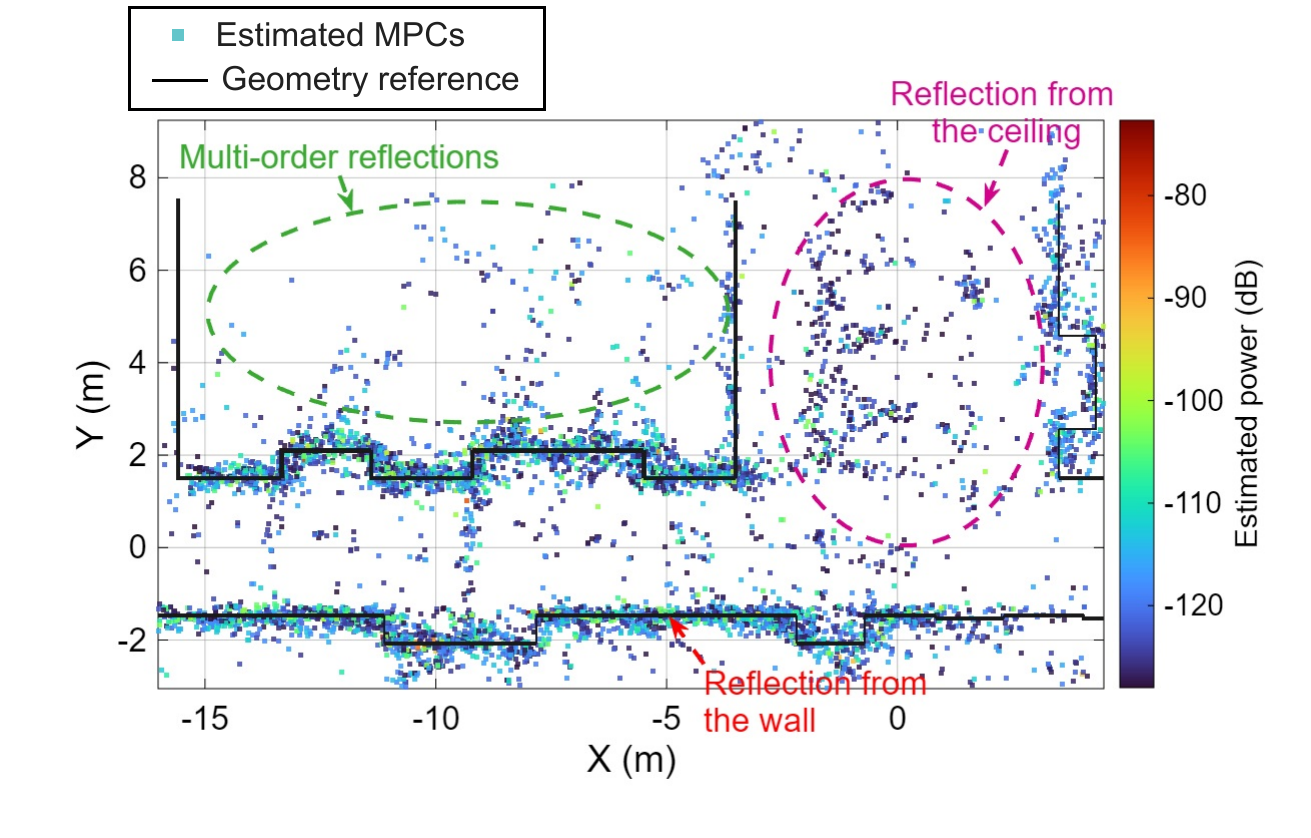}}
	\caption {The MPC mapping results compared with the geometry reference. (a) 3D view. (b) Top-view.}\label{fig:mapping}
\end{figure}

The Space-Alternating Generalized Expectation-Maximization (SAGE) algorithm is a widely used high-resolution parameter estimation technique for extracting MPCs in the delay and angular domains by iteratively separating their individual contributions~\cite{Fleury1999sage,Guangzheng2025}. Building on the trajectory-tracking-assisted SAGE framework proposed in~\cite{lyu2025hybrid}, this work extends the processing methodology to the three-dimensional delay--azimuth--elevation domain. Each antenna pointing pair $(\theta,\vartheta)$ is treated as a virtual array element, and an element-wise SAGE procedure is applied to estimate MPC delays and pattern-weighted complex amplitudes over the resulting two-dimensional angular lattice. To mitigate antenna-pattern-induced distortions and enforce spatial consistency, MPC trajectories are tracked across neighboring azimuth--elevation samples, and a representative component is selected for each trajectory based on maximum received power.

As a result, the de-embedded MPC parameter set at the $m^{\mathrm{th}}$ sensing location is obtained as
\begin{align}
\hat{\boldsymbol{\Theta}}^{(m)} =
\left[
\begin{matrix}
\hat{\alpha}_{1}^{(m)} & \cdots & \hat{\alpha}_{L_m}^{(m)} \\
\hat{\tau}_{1}^{(m)}   & \cdots & \hat{\tau}_{L_m}^{(m)} \\
\hat{\phi}_{1}^{(m)}   & \cdots & \hat{\phi}_{L_m}^{(m)} \\
\hat{\psi}_{1}^{(m)}   & \cdots & \hat{\psi}_{L_m}^{(m)}
\end{matrix}
\right],
\label{eq:3d_final_params}
\end{align}
where $L_m$ denotes the number of extracted MPCs, and each MPC is characterized by its complex amplitude, delay, azimuth, and elevation.

Based on the extracted MPC parameters, each THz echo is interpreted under a single-bounce specular reflection assumption, which is appropriate for dominant reflections from large-scale indoor structures at THz frequencies. Combining the estimated propagation delay with the estimated azimuth-elevation, the extracted MPCs are mapped from the signal domain into the geometric domain as candidate reflection points along rays emanating from the sensing location. As illustrated in Fig.~\ref{fig:mapping}, the resulting MPC distributions exhibit strong spatial consistency with the underlying hallway geometry, forming contiguous clusters associated with dominant structural elements such as walls, ceilings, and corner-induced multi-bounce regions. These results confirm that the proposed 3D SAGE-based processing provides a reliable geometric representation of the environment and serves as an effective front end for subsequent structure discovery and multi-modal integration.

\subsubsection{AI-Assisted Signal Processing}
To identify environmental structures from the mapped THz MPCs, an unsupervised structure extraction method is applied directly in the geometric domain defined by the estimated MPC parameters. The proposed method aims to discover structure-consistent MPC groups and to infer their physical interpretations by exploiting intrinsic geometric and propagation regularities in the measured THz signals, without relying on supervised semantic labels, predefined structural templates, or rule-based geometric inversion. From an AI perspective, the approach belongs to the class of unsupervised, data-driven structure discovery methods, in which environmental structures are treated as latent variables inferred through relational reasoning among MPCs.

Based on the extracted MPC set $\hat{\boldsymbol{\Theta}}^{(m)}$ at the $m^{\mathrm{th}}$ sensing location, a graph
$\mathcal{G}^{(m)}=(\mathcal{V}^{(m)},\mathcal{E}^{(m)})$
is constructed independently for each location to capture local geometric and propagation consistency among MPCs observed from a fixed viewpoint. The vertex set $\mathcal{V}^{(m)}=\{\ell\}$ contains all MPCs extracted at location $m$, while the edge set $\mathcal{E}^{(m)}$ encodes pairwise consistency relationships within the same location. To enforce locality and angular continuity, edges are established only between MPCs that are adjacent on the azimuth--elevation sampling lattice,
\begin{equation}
\mathcal{N}(\ell)
=
\left\{
\ell' \;\middle|\;
\big|\hat{\varphi}^{(m)}_{\ell'}-\hat{\varphi}^{(m)}_\ell\big|\le \Delta_\varphi,\;
\big|\hat{\phi}^{(m)}_{\ell'}-\hat{\phi}^{(m)}_\ell\big|\le \Delta_\phi
\right\},
\end{equation}
where $\Delta_\varphi$ and $\Delta_\phi$ are determined by the angular scanning resolution of the sensing system. For each neighboring MPC pair $(\ell,\ell')\in\mathcal{N}(\ell)$, an affinity weight is defined as
\begin{equation}
w^{(m)}_{\ell,\ell'}
=
\exp\!\left(
-\frac{\big(\hat{\tau}^{(m)}_\ell-\hat{\tau}^{(m)}_{\ell'}\big)^2}{\sigma_\tau^2}
-\frac{\big(\hat{\alpha}^{(m)}_{\ell}-\hat{\alpha}^{(m)}_{\ell'}\big)^2}{\sigma_\alpha^2}
\right),
\end{equation}
and an undirected edge $(\ell,\ell')$ is included in $\mathcal{E}^{(m)}$ if $w^{(m)}_{\ell,\ell'} \ge \eta$. This per-location graph construction enforces local angular continuity and smooth propagation behavior, such that MPCs associated with the same extended reflecting surface form elongated, connected structures, whereas isolated scatterers or noise-induced components tend to remain weakly connected.

Unsupervised MPC clustering is first achieved by extracting the connected
components of the integration graph $\mathcal{G}^{(m)} = \bigl(\mathcal{V}^{(m)}, \mathcal{E}^{(m)}\bigr)$, where each vertex $v_\ell \in \mathcal{V}^{(m)}$ corresponds to an estimated
MPC characterized by $(\hat{\tau}^{(m)}_\ell,\hat{\varphi}^{(m)}_\ell,
\hat{\phi}^{(m)}_\ell)$, and edges encode geometric affinity in the
delay--angle domain. The resulting structure-consistent MPC clusters are
given by
\begin{equation}
\big\{\mathcal{C}^{(m)}_k\big\}_{k=1}^{K_m}
=
\mathrm{ConnComp}\!\left(\mathcal{G}^{(m)}\right),
\end{equation}
with $\mathcal{C}^{(m)}_k \subset \mathcal{V}^{(m)}$.

Each MPC $\ell \in \mathcal{C}^{(m)}_k$ is mapped to a spatial point
$\mathbf{x}^{(m)}_\ell \in \mathbb{R}^3$ using its estimated delay and
direction,
\begin{equation}
\mathbf{x}^{(m)}_\ell
=
\frac{c\,\hat{\tau}^{(m)}_\ell}{2}\,
\hat{\mathbf{u}}\!\left(\hat{\varphi}^{(m)}_\ell,\hat{\phi}^{(m)}_\ell\right),
\end{equation}
where $\hat{\mathbf{u}}(\cdot)$ denotes the unit direction vector.

Due to geometric adjacency and corner-induced connectivity, a single MPC cluster
$\mathcal{C}^{(m)}_k$ may still contain spatial points originating from multiple
planar wall segments. To separate these contributions, a random sample consensus (RANSAC)-based plane
extraction procedure is applied to each dominant cluster, yielding a set of
planar inlier subsets $\{\mathcal{P}_j\}$. For each $\mathcal{P}_j$, a refined
planar wall model $(\mathbf{n}_j,d_j)$ is estimated via least-squares fitting over
all inliers.

Although all points in $\mathcal{P}_j$ lie on the same plane, they may form
multiple spatially disconnected regions due to occlusions, sparse sampling, or
corner-induced fragmentation. To obtain geometrically coherent wall segments,
each $\mathcal{P}_j$ is further partitioned based on Euclidean connectivity.
Specifically, two points $\mathbf{x}_i$ and $\mathbf{x}_k$ are considered
connected if
\begin{equation}
\|\mathbf{x}_i - \mathbf{x}_k\| \le d_{\mathrm{patch}},
\end{equation}
where $d_{\mathrm{patch}} = 0.08$ is the distance threshold to separate the wall segments, resulting in a set of planar wall patches $\big\{\mathcal{W}_q\big\}_{q=1}^{Q}$, where each $\mathcal{W}_q$ represents a spatially contiguous planar wall segment and $Q$ denotes the total number of extracted wall patches.

Due to incomplete sampling around wall intersections, a single physical wall may be fragmented into multiple parallel wall patches, while adjacent orthogonal walls may appear spatially disconnected. To explicitly recover the geometric connectivity induced by wall intersections, a bridging wall construction strategy is adopted, in which corner structures are implicitly modeled through the construction of virtual wall supports. Specifically, for two wall patches $\mathcal{W}_q$ and $\mathcal{W}_{q+1}$, a bridging wall is constructed if their plane normals satisfy an orthogonality constraint
\begin{equation}
\bigl|\mathbf{n}_q^\top \mathbf{n}_{q+1}\bigr| \le \varepsilon_{\perp},
\end{equation}
where $\varepsilon_{\perp}$ is an orthogonality tolerance accounting for plane-normal estimation errors. In this work, a $5^{\circ}$ angular tolerance is adopted, corresponding to $\varepsilon_{\perp} = \sin(5^{\circ})$. And if their minimum spatial separation satisfies
\begin{equation}
\min_{\mathbf{x}\in\mathcal{W}_q,\;\mathbf{y}\in\mathcal{W}_{q+1}}
\|\mathbf{x}-\mathbf{y}\|
\le d_{\mathrm{bridge}},
\end{equation}
where $d_{\mathrm{bridge}}=0.8$~m is the distance threshold for wall bridging. 

When both conditions are met, a virtual vertical bridging wall patch $\mathcal{W}^{\mathrm{bridge}}_{q\rightarrow q+1}$ is constructed between the two wall patches to explicitly represent the missing wall connectivity at the intersection region. This bridging wall serves as a geometric proxy for the corner structure and enables a topology-consistent wall representation without requiring explicit corner detection or classification.

As a result, the proposed geometry-driven method outputs a unified structural representation $\mathbf{W}=\{\mathcal{W}_q\}_{q=1}^{Q'}$, where $Q'$ is the total number of identified wall segments. The output $\mathbf{W}$ jointly encodes all
reconstructed planar wall elements and their corner-induced geometric
connectivity, directly inferred from the spatial geometry of the measured
MPCs.


\begin{figure}
	\centering
	\subfigure []{\includegraphics[width=1\columnwidth]{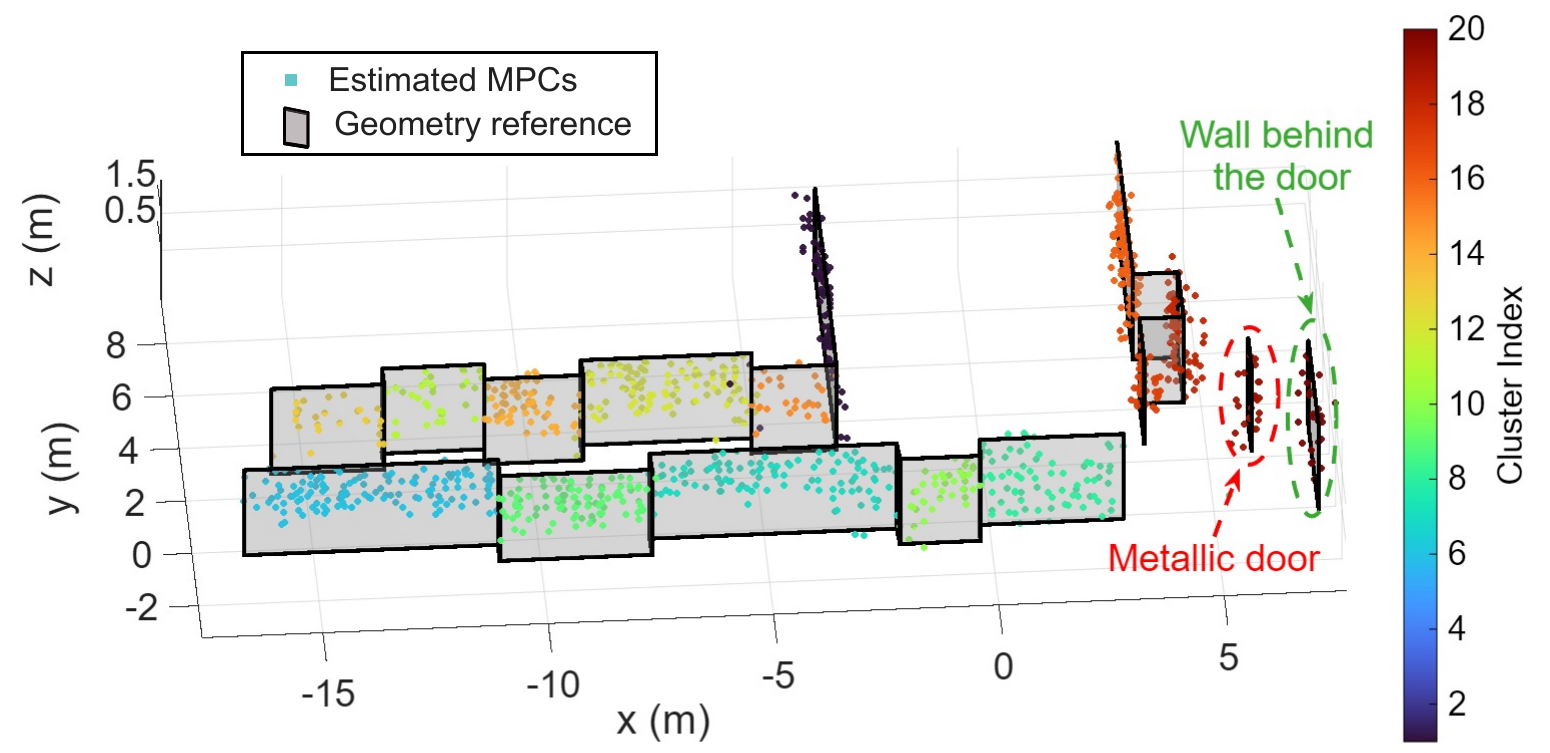}} 
	\subfigure []{\includegraphics[width=0.95\columnwidth]{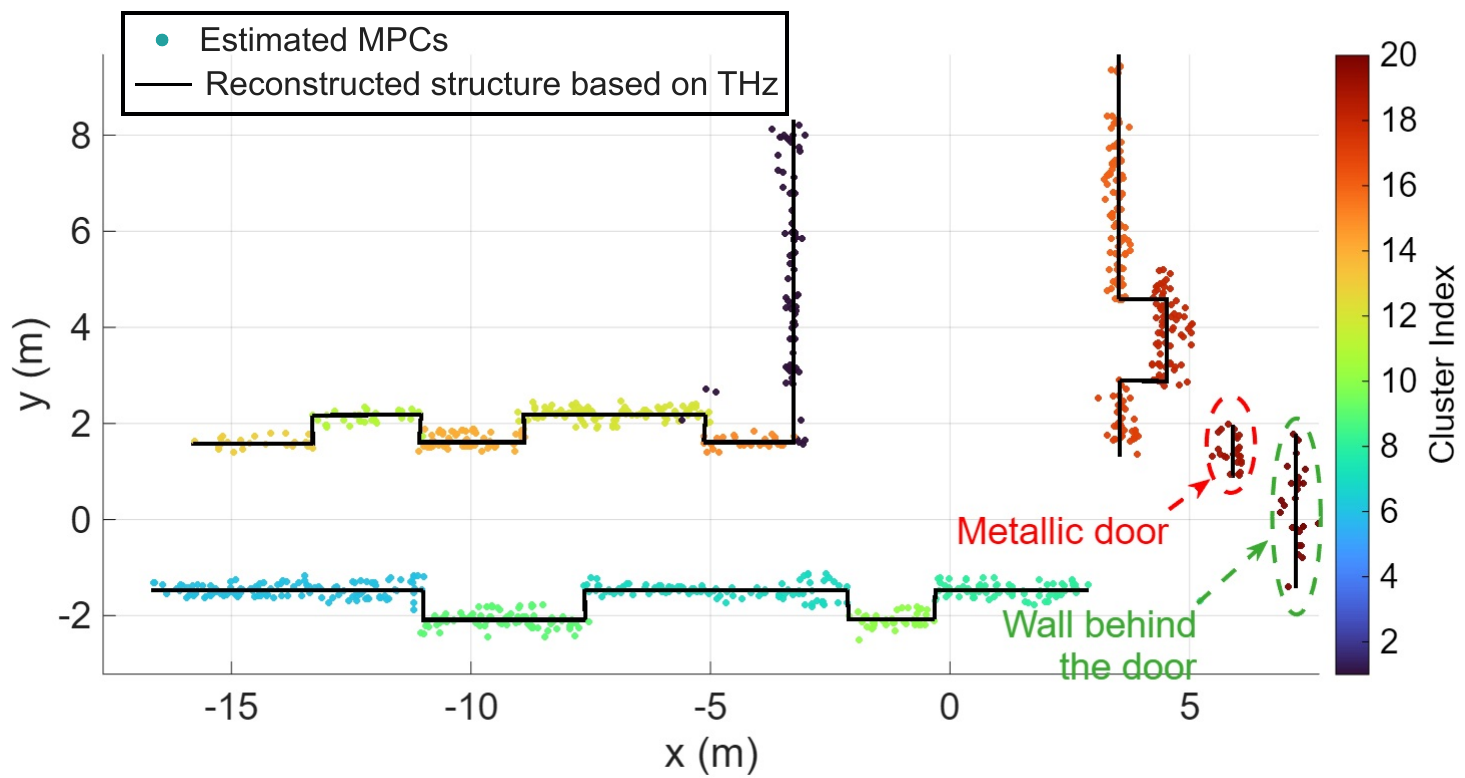}}
	\caption {Clustering results based on unsupervised graph-based integration and reconstructed wall structures.
(a) 3D view. (b) Top view.}\label{fig:thz_recon}
\end{figure}

\textbf{Material identification via reflection-loss analysis:}
After structural identification, material properties are inferred based on reflection-loss characteristics.
For the $\ell^{\mathrm{th}}$ MPC, the free-space path loss corresponding to $\hat{d}_{\ell}^{(m)}$ is
\begin{equation}
\mathrm{FSPL}(\hat{d}_{\ell}^{(m)}) =
\left( \frac{4\pi \hat{d}_{\ell}^{(m)}}{\lambda} \right)^2,
\end{equation}
where $\lambda$ is the carrier wavelength.
The residual attenuation is interpreted as reflection loss
\begin{equation}
\hat{L}_{\mathrm{refl},\ell}^{(m)}
=
-10\log_{10}
\left(
\frac{|\hat{\alpha}_{\ell}^{(m)}|^2}
{\mathrm{FSPL}(\hat{d}_{\ell}^{(m)})}
\right).
\end{equation}

Cluster-level reflection-loss statistics are then compared with the THz material database~\cite{fang2025environment} to infer the most plausible surface material associated with each identified structure.

Fig.~\ref{fig:thz_recon} presents the clustering and geometry-based reconstruction results obtained using the proposed method. The reduced number of visualized MPCs compared with Fig.~\ref{fig:mapping} is solely due to visualization-level selection for clarity and does not result from the RANSAC-based plane extraction or the unsupervised clustering process. The extracted MPCs are partitioned into $20$ clusters, each corresponding to a structure-consistent group of propagation paths that captures the geometric characteristics of a dominant environmental structure. As observed, the reconstructed wall surfaces are well aligned with the spatial extent and orientation of their associated MPC clusters, validating the effectiveness of the proposed clustering and geometry-driven reconstruction framework. Owing to unfavorable sensing geometry and weak received signal strength in certain directions, some wall regions are only sparsely illuminated by the THz measurements, leading to locally discontinuous reconstruction areas, as highlighted in Fig.~\ref{fig:thz_recon}. In addition, a metallic door introduces a structural discontinuity in the reconstructed wall. Notably, the wall region behind the door is still partially detected by the THz sensing system, even though it is not visible in the fisheye camera view. This observation directly demonstrates the non-line-of-sight (NLOS) sensing capability of monostatic THz measurements, highlighting a key advantage over purely vision-based sensing modalities. Due to the fact that each THz cluster may comprise a mixture of multiple specular and diffuse reflection components that blur structural boundaries and continuity, which leads to a unseparation of the cluster and material while using only THz model, complementary visible-light sensing is incorporated to provide global shape cues and continuity constraints, enabling reliable verification and refinement of the reconstructed wall geometries.

\begin{figure}
	\centering
	\subfigure []{\includegraphics[width=0.97\columnwidth]{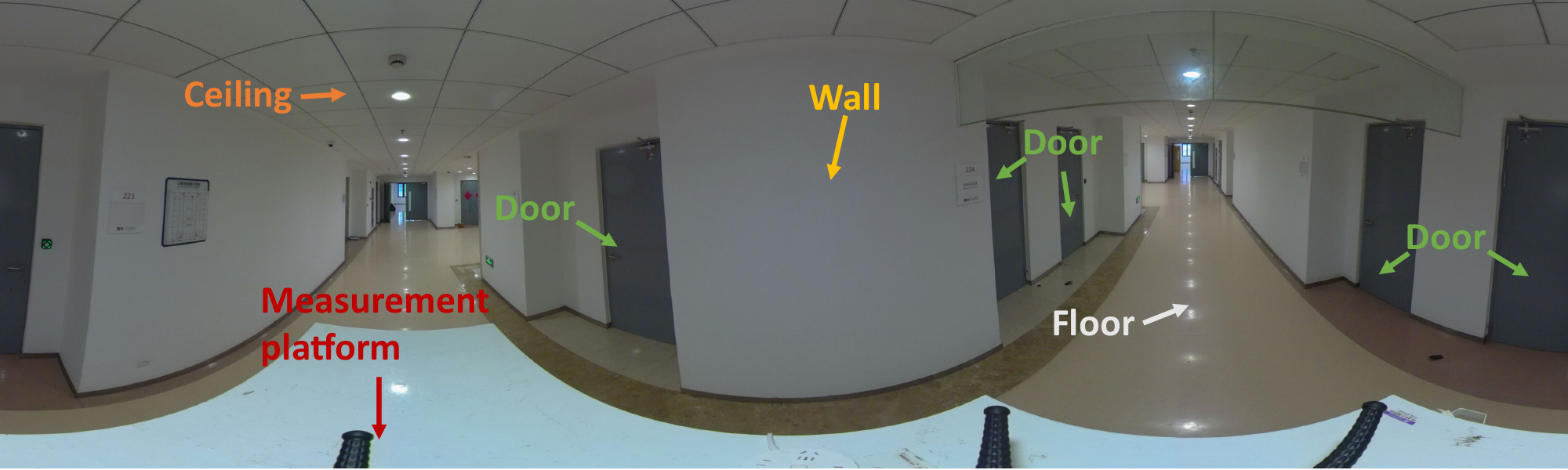}} 
	\subfigure []{\includegraphics[width=0.92\columnwidth]{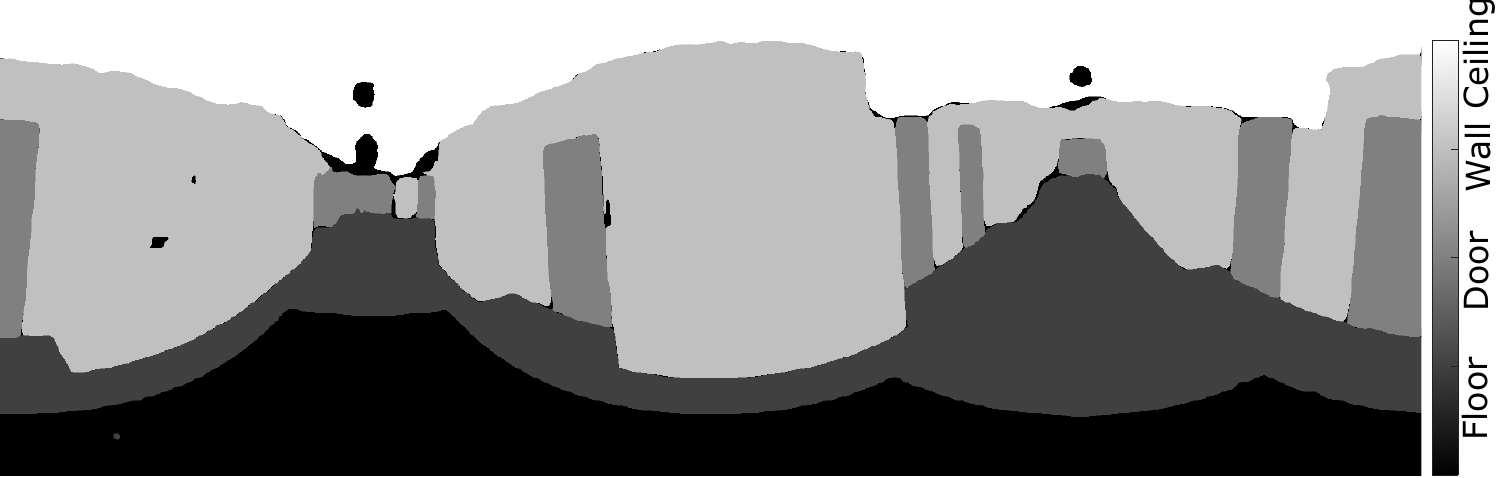}}
    \subfigure []{\includegraphics[width=1\columnwidth]{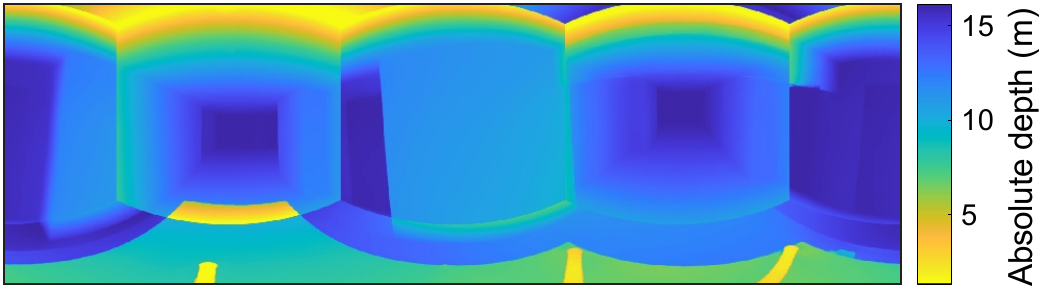}}
	\caption {The example of the panoramic image, semantic mask map, and estimated depth map. (a) Panoramic image in the Location~$12$. (b) Semantic mask map ($K=5$). (c) Estimated depth map.}\label{fig:semantic_example}
\end{figure}

\subsection{AI-Based Methods for Visible-Light}
\label{subsec:segmenter_panorama}
\subsubsection{Semantic Perception}
To extract semantic information from the visible-light modality and to enable subsequent multi-modal integration with THz sensing results, an AI-based semantic perception method is developed for panoramic images captured by a fisheye camera. The panoramic image provides an omnidirectional representation of the measurement scenario, which is particularly suitable for establishing correspondence with angular-domain THz trajectory-tracked MPCs. In this work, the term ``semantic'' refers to category-level environmental meaning assigned to visually observed regions, rather than only raw object names or low-level image features. Specifically, the semantic categories considered in this work, such as wall, floor, ceiling, door, and metallic object, describe the physical or functional roles of different scene regions in the reconstructed environment.

After geometric calibration and projection normalization, each pixel in the panoramic image is mapped to a unique angular direction characterized by azimuth $\phi \in [-\pi,\pi)$ and elevation $\varphi \in [-\pi/2,\pi/2]$. The resulting panoramic representation is denoted as $\mathcal{I}(\phi,\varphi)$, providing a direct angular-domain interface for multi-modal integration. Semantic segmentation is then performed using a transformer-based method following the Segmenter paradigm~\cite{strudel2021segmenter}. Segmenter is a fully transformer-based semantic segmentation architecture, consisting of a Vision Transformer encoder for patch-level feature extraction and a lightweight decoder for pixel-level mask prediction. In our implementation, the panoramic images are processed using a Segmenter model with ADE$20$K dataset. The segmentation outputs are then converted into angular-domain semantic mask maps and associated with THz-derived structural components in the subsequent cross-modal integration stage.Semantic segmentation is adopted since the proposed cross-modal integration module requires both category labels and angular support regions. Unlike image classification or bounding-box-based detection, segmentation provides dense mask-level boundaries that can be directly converted into angular-domain representations and associated with THz-derived MPCs through angular-overlap analysis.

The semantic segmentation network produces a dense angular-domain semantic representation of the observed scene. For each angular direction $(\phi,\varphi)$ on the panoramic sphere, the network outputs both a semantic class label and an associated confidence score over a predefined set of semantic categories, including walls, floors, ceilings, doors, glass surfaces, and metallic objects. Let $s(\phi,\varphi) \in \{1,2,\ldots,K\},$ denote the predicted semantic class index at direction $(\phi,\varphi)$, where $K$ is the total number of semantic categories, and let $p(\phi,\varphi)\in[0,1]$ denote the corresponding confidence derived from the network’s softmax output. This angular-domain semantic representation encodes not only class identity but also the angular shape and separation of different semantic regions on the unit sphere. As such, it provides a structured semantic abstraction of the visual scene and establishes a natural interface for subsequent cross-modal integration with THz-derived geometric structures.


Beyond pixel-level semantic labeling, the agent further analyzes the angular-domain semantic map to extract region-level information. By aggregating semantic labels over connected angular regions, spatially coherent semantic masks are identified, corresponding to dominant environmental structures. These semantic masks encode both the class information (e.g., wall, door, metallic object) and the geometric shape information (e.g., angular extent and orientation) of major objects in the scene. Fig.~\ref{fig:semantic_example} presents a representative example of the panoramic image and the corresponding semantic mask map at Location~11. Owing to the wide field-of-view provided by the fisheye camera, an omnidirectional $360^{\circ}$ panoramic image of the measurement environment is captured, as shown in Fig.~\ref{fig:semantic_example}~(a), enabling comprehensive visual coverage of the surrounding structures.

The semantic mask map, illustrated in Fig.~\ref{fig:semantic_example}~(b), assigns a semantic label to each pixel in the panoramic image, resulting in a dense, pixel-wise representation of the scene. In this example, four dominant object classes and geometric structures are identified, namely the floor, doors, walls, and ceiling. The extracted semantic regions exhibit clear boundaries and strong spatial consistency, accurately reflecting the physical layout of the indoor hallway scenario. In particular, large planar surfaces such as walls and ceilings form continuous semantic regions, while doors appear as compact, high-contrast semantic entities embedded within the wall structures.

The resulting semantic mask map provides explicit object-level class and shape information that is difficult to infer directly from THz sensing measurements alone. This semantic representation serves as an important prior for subsequent channel modeling and environment reconstruction. By associating the THz sensing-derived multipath components with their corresponding semantic classes and angular extents, material-dependent scattering behavior and reflection mechanisms can be more reliably interpreted. Furthermore, the panoramic semantic representation establishes a direct correspondence between angular-domain THz sensing results and the physical scene geometry, enabling semantic-aware channel modeling and geometry-consistent environment reconstruction.

\subsubsection{Depth Estimation}
In addition to semantic segmentation, an AI-based depth estimation module is employed to extract dense vision-based depth map from the panoramic image. Since a single fisheye image cannot directly provide physically measured metric distance, the estimated depth is not treated as a direct range measurement, but as an auxiliary geometric prior for subsequent multi-modal integration. In this work, Depth Anything is adopted to estimate monocular depth from visible-light images~\cite{depth_anything}.

Directly applying monocular depth estimation to a full equirectangular panorama may introduce distortion-induced depth inconsistency, since the panoramic image does not follow a standard perspective projection. To alleviate this issue, the panoramic image is first divided into several overlapping perspective sub-views along the azimuth direction. Each sub-view has a limited field of view and better approximates the input format of a standard monocular depth estimation model. Depth estimation is then performed independently for each perspective sub-view. The estimated depth maps are subsequently re-projected to the panoramic coordinate system and stitched to obtain a complete panoramic depth map.

Since monocular depth estimation mainly provides relative depth, sparse reference range information from light detection and ranging (LiDAR) measurements is used to calibrate the absolute scale of the estimated depth map. The calibrated depth map is then compared with the THz-derived range and structural information to check geometric consistency. As illustrated in Fig.~\ref{fig:semantic_example}~(c), the estimated depth map reflects the corridor extension, side-wall distribution, and major structural layout of the indoor scenario, while also preserving the relative depth consistency among different scene components, e.g., wall. Nevertheless, local inconsistencies may still appear around corners, wall-ceiling boundaries, and stitching regions between adjacent perspective sub-views, due to projection distortion, scale-calibration uncertainty, and the limited ability of RGB-based depth estimation to distinguish geometrically similar indoor surfaces. Therefore, the vision-derived depth map is used as a reliability-dependent geometric prior, while the THz-derived range and structural information remain the primary physical references for metric reconstruction.

\begin{figure*}
	\centering
	\includegraphics[width=1.6\columnwidth]{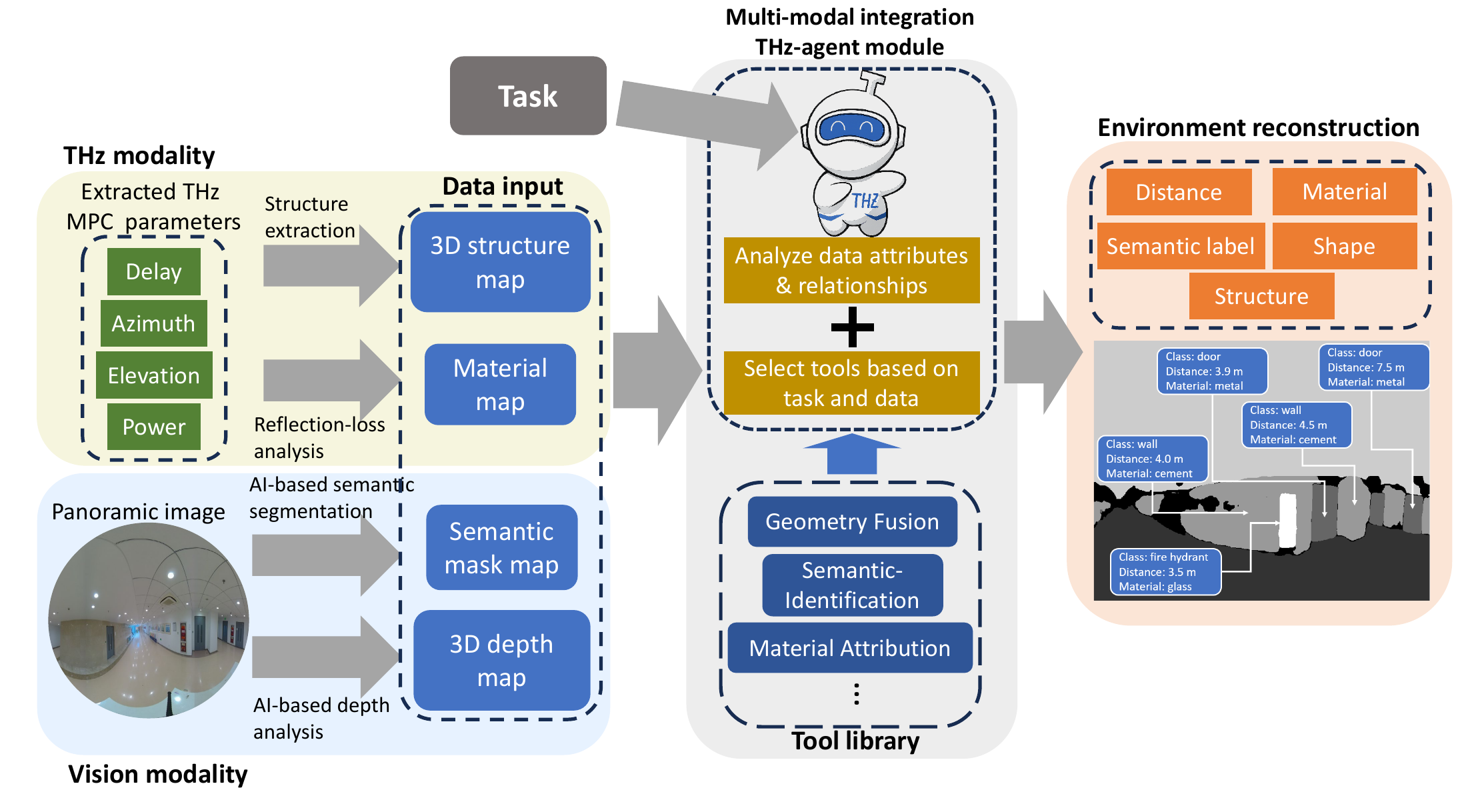}
	\caption {\textcolor{blue}The THz-agent framework for the multi-modal environment reconstruction.}\label{fig:agent_framework}
\end{figure*}

\subsection{Discussion on the Capability of the Two Modalities}
The THz modality is a physics-driven sensing modality that observes the environment through delay-, angle-, and power-resolved electromagnetic propagation responses. Benefiting from the ultra-wide bandwidth and directional scanning mechanism, THz sensing can provide range-aware structural information and can localize dominant reflectors with high metric accuracy. In addition, reflection-loss characteristics further provide material-related cues, which are useful for interpreting the physical properties of reconstructed structures. Therefore, the THz modality mainly serves as the geometric, metric, and material anchor in the proposed framework. However, THz sensing is also sparse and reflection-dependent. Its observations may become discontinuous in regions with weak backscatter, unfavorable incidence angles, or limited angular coverage. Moreover, THz measurements do not directly encode object-level semantic meanings, and different structures with similar geometry or material responses may produce comparable propagation signatures.

In contrast, the vision modality provides dense appearance-domain observations over the panoramic field of view. Through semantic segmentation, the visible-light image provides object-level class labels, angular support regions, and boundary information for major scene components such as walls, floors, ceilings, doors, and other indoor objects. Thus, the vision modality mainly serves as the semantic and contextual anchor in the proposed framework. Nevertheless, in the considered single-camera panoramic setup, vision does not directly provide reliable metric distance information, and its perception performance can be affected by illumination variation, texture ambiguity, and occlusion. Therefore, neither modality alone is sufficient for geometry-consistent and semantically meaningful environment reconstruction. By aligning the two modalities in a common angular domain, the proposed framework exploits THz sensing to constrain metric geometry and material attributes, while using vision sensing to provide dense semantic interpretation and angular object boundaries. This complementarity enables the final reconstruction to be both physically consistent and semantically interpretable.

\section{Multi-Modal Environment Reconstruction}\label{sec:material}
This section presents the multi-modal environment reconstruction process, including angular alignment, THz-agent-based integration, and reconstruction evaluation. Before integration, the THz sensing results and panoramic visual observations are aligned into a unified angular reference.

\subsection{Task-Driven THz-Agent Module for Multi-Modal Reconstruction}
After angular alignment, the outputs from the THz and vision modalities are integrated by a task-driven THz-agent module, as illustrated in Fig.~\ref{fig:agent_framework}. Different from a fixed sequential fusion pipeline, the THz-agent acts as an expert-based controller that analyzes the attributes of the available modality-specific information, selects suitable tools from a predefined tool library according to the reconstruction task, and generates a unified environment representation. The aim of the THz-agent is not to replace the physical sensing models, but to organize THz-derived metric geometry and material information, vision-derived semantic information, and vision-derived depth priors in a task-conditioned manner. In this work, the target task is geometry-consistent and semantically interpretable indoor environment reconstruction.

The THz-agent takes four modality-specific inputs, including the THz-derived structural map $\mathcal{G}_{\mathrm{THz}}$, the THz-derived material map $\mathcal{M}_{\mathrm{THz}}$, the vision-derived semantic mask map $\mathcal{Y}_{\mathrm{vis}}$, and the calibrated vision-derived depth map $\mathcal{D}_{\mathrm{vis}}$. The overall input set is expressed as
\begin{equation}
\mathcal{X}
=
\{\mathcal{G}_{\mathrm{THz}},
\mathcal{M}_{\mathrm{THz}},
\mathcal{Y}_{\mathrm{vis}},
\mathcal{D}_{\mathrm{vis}}\}.
\end{equation}
Here, $\mathcal{G}_{\mathrm{THz}}=\{W_q\}_{q=1}^{Q}$ represents the THz-derived structural elements, $\mathcal{M}_{\mathrm{THz}}=\{\bar{L}_{\mathrm{refl},q},\hat{m}^{\mathrm{THz}}_q\}_{q=1}^{Q}$ contains the reflection-loss statistics and material estimates from THz measurements, $\mathcal{Y}_{\mathrm{vis}}=\{s(\varphi,\psi),p(\varphi,\psi),\Omega_k\}_{k=1}^{K}$ provides the angular semantic labels $s(\varphi,\psi)$, confidence scores $p(\varphi,\psi)$, and semantic support regions $\Omega_k$, and $\mathcal{D}_{\mathrm{vis}}=\{\hat{D}_{\mathrm{vis}}(\varphi,\psi)\}$ denotes the calibrated vision-derived depth prior. Among these inputs, the THz-derived structural and material maps provide the primary metric and physical references, while the vision-derived semantic and depth maps provide auxiliary semantic and dense geometric priors.

Given the reconstruction task $\mathcal{T}$ and the multi-modal input set $\mathcal{X}$, the THz-agent analyzes whether each input mainly contributes geometry, material, semantic, or depth-related information, and then selects task-relevant tools from the tool library. The tool selection process is expressed as
\begin{equation}
\mathcal{A}_{\mathcal{T}}
=
\Gamma(\mathcal{T},\mathcal{X}),
\end{equation}
where $\Gamma(\cdot)$ denotes the task-driven tool-selection function and $\mathcal{A}_{\mathcal{T}}$ is the selected set of tools for the current reconstruction task. Several representative tools are described below.

The first example is the geometry-fusion tool, which converts the calibrated vision-derived depth map into a 3D representation and combines it with the THz-derived planar structural geometry using a reliability-dependent weight. For the $q$-th THz-derived planar structure $W_q$, the vision-derived depth values within its corresponding angular support are first back-projected into 3D space and fitted as a vision-supported planar component. Let $\boldsymbol{\theta}_{\mathrm{THz},q}=[\mathbf{n}_q^{\mathrm{T}},d_q]^{\mathrm{T}}$ denote the THz-derived plane parameter, and let $\boldsymbol{\theta}_{\mathrm{vis},q}=[\mathbf{n}_{\mathrm{vis},q}^{\mathrm{T}},d_{\mathrm{vis},q}]^{\mathrm{T}}$ denote the corresponding plane parameter estimated from the vision-derived depth map. The fused planar geometry is obtained by
\begin{equation}
\boldsymbol{\theta}_{\mathrm{re},q}
=
\mathcal{N}\left(
\omega_q \boldsymbol{\theta}_{\mathrm{THz},q}
+
(1-\omega_q)\boldsymbol{\theta}_{\mathrm{vis},q}
\right),
\end{equation}
where $\omega_q\in[0,1]$ is the reliability-dependent weight assigned to the THz-derived geometry, and $\mathcal{N}(\cdot)$ denotes normalization of the fused plane normal. In this work, the THz-derived structural geometry serves as the primary metric reference, while the vision-derived depth provides an auxiliary dense geometric prior to refine the spatial extent and consistency of the reconstructed planar structure. The fused planar components form the reconstructed geometry $\mathcal{G}_{\mathrm{re}}=\{G_{\mathrm{re},q}\}_{q=1}^{Q'}$.

Second, the semantic-identification tool identifies object categories and separates object-level shapes on the reconstructed geometry using the vision-derived angular semantic masks. Specifically, the semantic mask map provides the angular support region of each semantic category, such as wall, floor, ceiling, door, or local object, in the panoramic field of view. For a reconstructed geometric component $G_{\mathrm{re},q}$, each spatial point is projected from the TRx location to the panoramic angular domain and compared with the semantic support regions. In this way, the angular shape of each semantic label is back-projected onto the reconstructed geometry, enabling semantic identification and separation of different components, such as doors from wall regions or local objects from permanent structural surfaces. The assigned semantic label of $G_{\mathrm{re},q}$ is denoted by the semantic label $\hat{s}_q$.

Third, the material-attribution tool directly attaches the material information inferred from THz reflection-loss analysis to the reconstructed geometry,
\begin{equation}
\widetilde{G}_{\mathrm{re},q}
=
\{G_{\mathrm{re},q},\hat{m}^{\mathrm{THz}}_q,\bar{L}_{\mathrm{refl},q}\},
\end{equation}
where $\hat{m}^{\mathrm{THz}}_q$ denotes the material category inferred from the THz measurements and $\bar{L}_{\mathrm{refl},q}$ is the structure-level reflection-loss statistic.

Through the selected tools, the THz-agent generates a unified environment representation
\begin{equation}
\mathcal{E}_{\mathrm{re}}
=
\{G_{\mathrm{re},q},\hat{s}_q,\hat{m}_q,\rho_q\}_{q=1}^{Q'},
\end{equation}
where $\rho_q$ is the reliability score determined by the selected geometry, semantic, depth, and material consistency tools. Note that the tool library is not limited to the representative tools described above. It can be further extended to include additional tools for multi-modal data integration when more sensing modalities are available, such as millimeter-wave sensing, LiDAR, or other visual sensors.

\begin{figure}
	\centering
	\includegraphics[width=1\columnwidth]{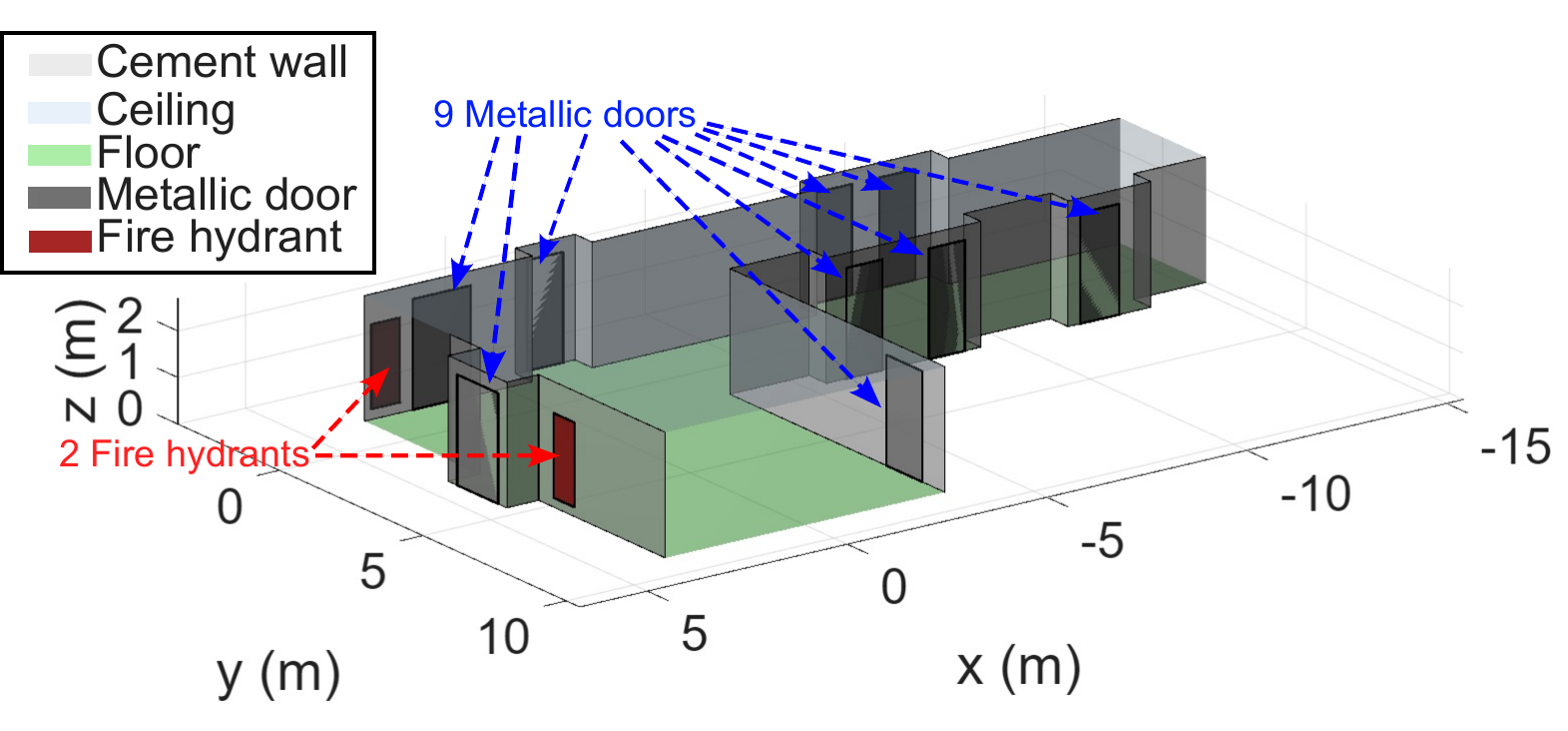}
	\caption {The reconstructed environment result.}\label{fig:recon_envir}
\end{figure}

\begin{figure}
	\centering
	\subfigure []{\includegraphics[width=1\columnwidth]{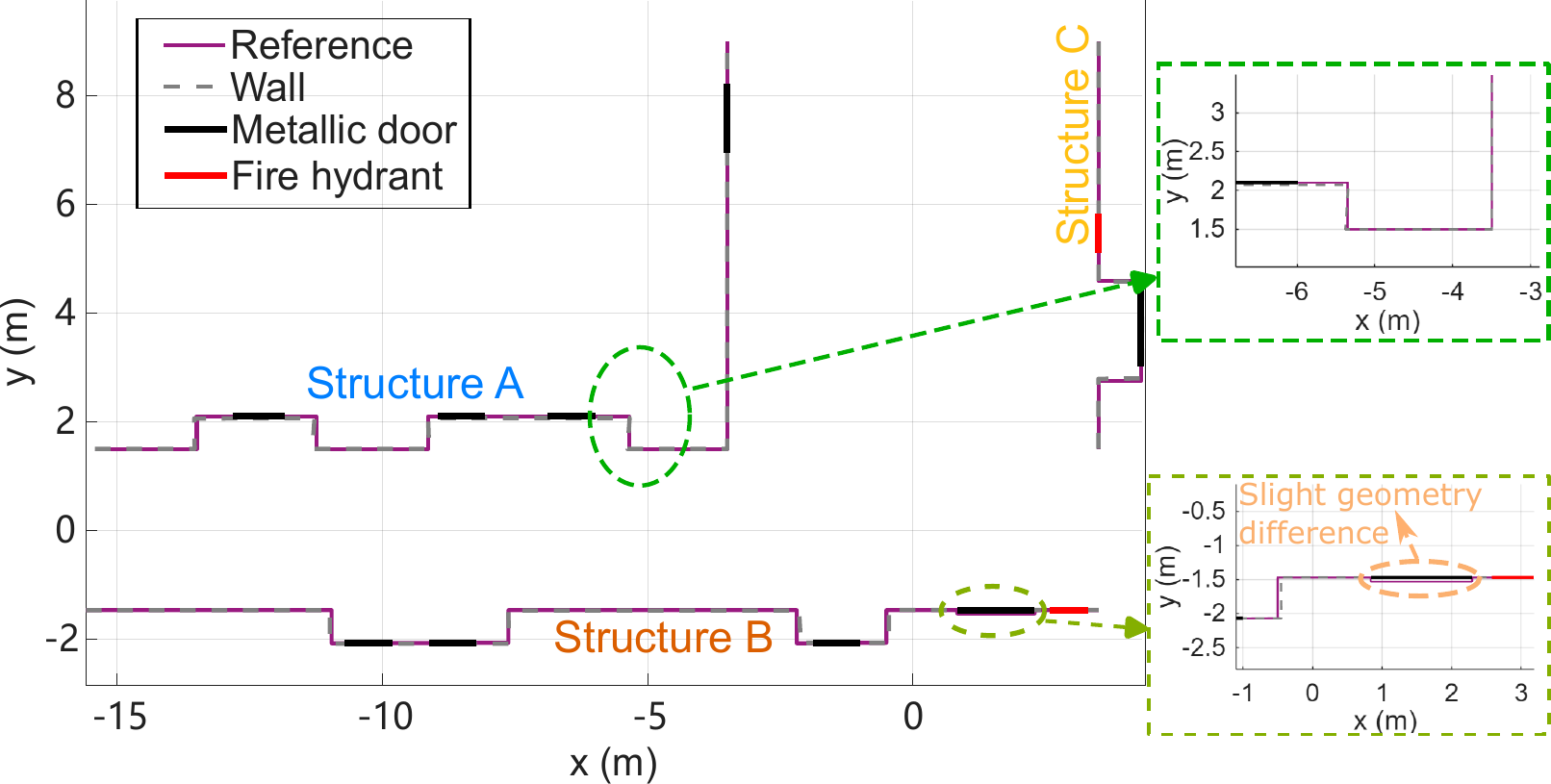}} 
	\subfigure []{\includegraphics[width=0.85\columnwidth]{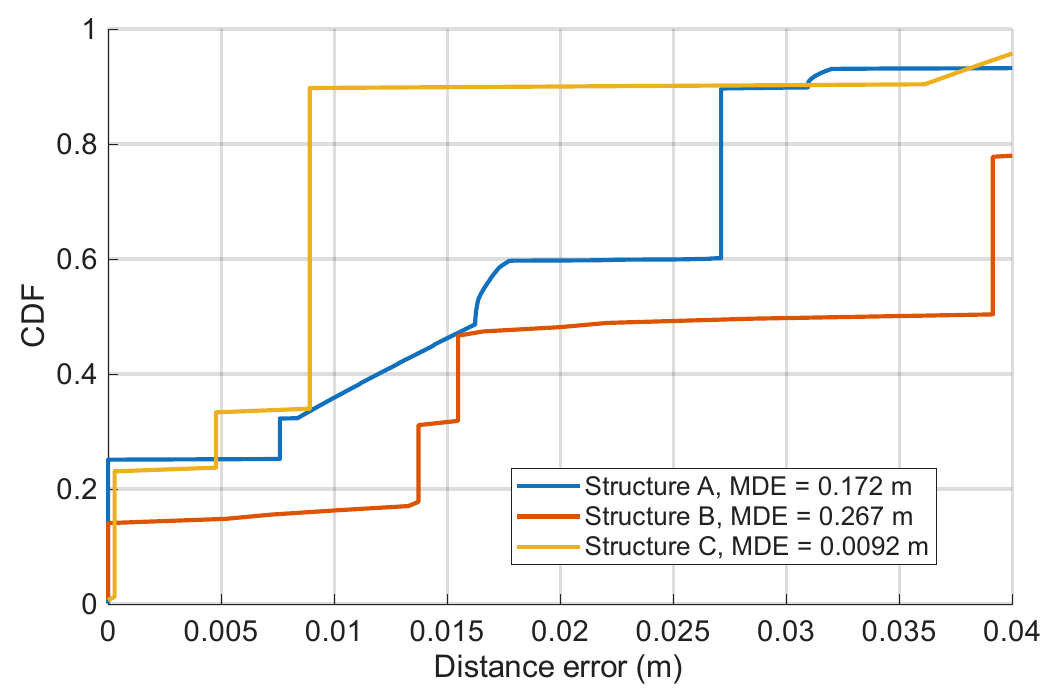}}
	\caption {Environment reconstruction in 2D view and error analysis.
(a) Geometric comparison between the reconstructed environment and the reference layout in the 2D plane. (b) CDF of the reconstruction error with respect to the reference.}\label{fig:error}
\end{figure}

\subsection{Reconstruction Results}
Based on the proposed framework, 3D environment reconstruction is achieved by jointly leveraging multi-modal sensing information in an angle-conditioned manner. The reconstruction process integrates semantic and geometric cues inferred from the two modalities to produce a spatially consistent and semantically interpretable representation of the environment.

The reconstructed indoor L-shaped hallway scenario is illustrated in Fig.~\ref{fig:recon_envir}, where the proposed multi-modal reconstruction is compared with the corresponding single-modal results, namely the THz-based reconstruction in Fig.~\ref{fig:thz_recon} and the visible-light-based reconstruction in Fig.~\ref{fig:semantic_example}. As shown in the THz-only result in Fig.~\ref{fig:thz_recon}, the dominant wall trajectories and the overall hallway layout can be recovered by fitting geometry-consistent MPC clusters. However, the reconstructed structures exhibit noticeable discontinuities in certain regions due to unfavorable geometry and weak received signals, which, together with the absence of semantic information, limit the reliable identification of doors and smaller objects. In contrast, the visible-light-only result provides dense semantic labeling with clear class boundaries for walls, floor, ceiling, and doors in the angular domain, but remains inherently view-dependent and does not yield accurate metric depth or a consistent 3D geometry, as illustrated in Fig.~\ref{fig:semantic_example}. By comparison, the multi-modal reconstruction obtained using the proposed framework effectively integrates the complementary strengths of both modalities. The cement walls and ceiling are reconstructed as continuous planar surfaces with consistent spatial alignment, while the floor region is accurately delineated with correct geometric extent. Metallic doors are clearly identified at their corresponding locations, exhibiting coherent planar geometry and consistent material attribution, and the fire hydrant is reconstructed as a compact object with correct semantic classification, material identification, and spatial placement. The multi-modal result captures the global hallway geometry, including wall continuity, door recesses, and structural transitions along the hallway, achieving a level of completeness and interpretability that cannot be attained by either sensing modality alone.

To quantitatively evaluate the reconstruction fidelity, the reconstructed geometry is compared with the reference ground truth, as illustrated in Fig.~\ref{fig:error}. As shown in Fig.~\ref{fig:error}~(a), the reconstructed environment exhibits close geometric alignment with the reference layout in the $2$D plane, indicating accurate recovery of the dominant structural features. Fig.~\ref{fig:error}~(b) further presents the cumulative distribution function (CDF) of the distance error. The mean distance errors for Structure~A, Structure~B, and Structure~C are $17.2$, $26.7$, and $9.3$~mm, respectively, demonstrating the effectiveness of the proposed framework in enabling high-precision environment reconstruction. Among the evaluated structures, Structure~C achieves the highest reconstruction accuracy, whereas Structure~B exhibits the largest error. This degradation is primarily attributed to the incomplete reconstruction of a localized but very slight geometric variation, with a characteristic scale of approximately $0.06$~m, as highlighted in the bottom-right subfigure of Fig.~\ref{fig:error}~(a). Although minor in magnitude, such small-scale structural deviations can be insufficiently captured under insufficient angular sampling conditions. Addressing this limitation constitutes an important direction for future work. Overall, these results confirm that the proposed environment reconstruction framework can achieve high spatial fidelity while maintaining geometry-consistent structural interpretation under the considered measurement conditions. Such performance is essential for downstream applications, including digital-twin construction and environment-aware wireless system design.

Although the present measurement campaign focuses on an indoor L-shaped hallway scenario without intentionally placed obstacles, the proposed framework is not restricted to this specific layout. In principle, it can be extended to more complex environments, such as dense-reflection indoor scenes, factories, warehouses, and laboratories, since its core components are based on general sensing principles: THz-based structural inference from propagation measurements and vision-based semantic perception from panoramic images. In such environments, THz sensing can still provide physically grounded geometric cues from dominant reflections, while vision sensing can provide dense semantic and contextual information for surrounding objects, machinery, and structural elements. Nevertheless, more complex scenes also introduce additional challenges. Metallic surfaces, cluttered layouts, irregular machinery, boxes, and partial occlusions may increase the number of MPCs, causing stronger multipath overlap, diffuse scattering, shadowed angular sectors, and more ambiguous structural hypotheses. For example, an obstacle placed in the corridor may generate additional localized MPCs while blocking or attenuating the background wall or floor returns, leading to incomplete spatial extents or reduced structural continuity mainly in the obstacle-shadowed regions. Compared with THz-only reconstruction, the proposed multi-modal framework can partly mitigate this issue by using vision-derived semantic masks to distinguish local obstacles from permanent structural elements, while THz sensing provides range-aware geometric information. However, if a region is fully occluded in both THz and vision observations, it cannot be reliably recovered from a single monostatic viewpoint alone. Therefore, applying the proposed framework to such scenarios may require further extensions, including multi-view measurements, adaptive angular scanning, advanced multipath clustering and separation, reliability-aware cross-modal association, multi-bounce propagation modeling, and richer geometric representations beyond planar wall primitives. These extensions constitute important directions for future work toward robust multi-modal THz sensing in general complex environments.

\section{Conclusion}\label{sec:conclusion}
This paper presents a measurement-based multi-modal sensing and environment reconstruction framework that integrates monostatic THz sensing with visible-light perception. The proposed sensing architecture jointly deploys a 3D delay-azimuth-elevation THz monostatic channel sounder and an omnidirectional fisheye camera, enabling the geometry-consistent acquisition and angular alignment of radio-frequency and visual observations from a common sensing viewpoint. Based on the measured THz data, a model-driven signal processing pipeline is developed to extract spatially consistent MPCs and infer geometry- and material-consistent structural primitives through trajectory-tracking-based parameter estimation, graph-based structure discovery, planar reconstruction, and reflection-loss analysis. In parallel, AI-based vision modules are employed to extract object-level semantic categories, angular support regions, and depth map from panoramic imagery. To integrate these heterogeneous modality-specific outputs, an agentic-AI-based task-driven THz-agent framework is proposed, in which angle-conditioned alignment and consistency analysis associate THz-derived metric geometry and material information with vision-derived semantic regions and depth maps. Experimental validation in the indoor L-shaped hallway scenario demonstrates that the proposed framework achieves centimeter-level geometric accuracy, with locally millimeter-level accuracy for selected structures, while identifying dominant structural elements and representative indoor objects. These results demonstrate the effectiveness of measurement-grounded THz–vision integration for physically interpretable environment reconstruction, and provide a foundation for future environment-aware THz sensing, ISAC systems, and digital-twin-enabled wireless networks.

\normalem
\bibliography{monostatic_sensing}

@ARTICLE{isac_ris,
  author={Ma, Zhangfeng and Liang, Yongzhe and Zhu, Qiuming and Zheng, Jiakang and Lian, Zhuxian and Zeng, Linzhou and Fu, Chengwei and Peng, Yifei and Ai, Bo},
  journal={IEEE Trans. Cogn. Commun. Netw.}, 
  title={Hybrid-{RIS}-Assisted Cellular {ISAC} Networks for {UAV}-Enabled Low-Altitude Economy via Deep Reinforcement Learning with Mixture-of-Experts}, 
  year={2025},
  volume={},
  number={},
  pages={1-1},
  doi={10.1109/TCCN.2025.3622130}}

@article{cheng2026apeg,
  title={{APEG}: Adaptive Physical Layer Authentication with Channel Extrapolation and Generative {AI}},
  author={Cheng, Xiqi and Meng, Rui and Xu, Xiaodong and Gao, Haixiao and Zhang, Ping and Niyato, Dusit},
  journal={IEEE Transactions on Information Forensics and Security},
  year={2026},
  volume={21},
  pages={1257-1272},
doi={10.1109/TIFS.2026.3654380}
}

@INPROCEEDINGS{depth_anything,
  author={Yang, Lihe and Kang, Bingyi and Huang, Zilong and Xu, Xiaogang and Feng, Jiashi and Zhao, Hengshuang},
  booktitle={2024 IEEE/CVF Conference on Computer Vision and Pattern Recognition (CVPR)}, 
  title={Depth Anything: Unleashing the Power of Large-Scale Unlabeled Data}, 
  year={2024},
  volume={},
  number={},
  pages={10371-10381},
  doi={10.1109/CVPR52733.2024.00987}}

@ARTICLE{6G_isac1,
  author={Zhang, Zhengquan and Xiao, Yue and Ma, Zheng and Xiao, Ming and Ding, Zhiguo and Lei, Xianfu and Karagiannidis, George K. and Fan, Pingzhi},
  journal={IEEE Veh. Technol. Mag.}, 
  title={{6G} Wireless Networks: Vision, Requirements, Architecture, and Key Technologies}, 
  year={2019},
  volume={14},
  number={3},
  pages={28-41},
  doi={10.1109/MVT.2019.2921208}}

@ARTICLE{6G_isac2,
  author={Wang, Cheng-Xiang and You, Xiaohu and Gao, Xiqi and Zhu, Xiuming and Li, Zixin and Zhang, Chuan and Wang, Haiming and Huang, Yongming and Chen, Yunfei and Haas, Harald and Thompson, John S. and Larsson, Erik G. and Renzo, Marco Di and Tong, Wen and Zhu, Peiying and Shen, Xuemin and Poor, H. Vincent and Hanzo, Lajos},
  journal={IEEE Commun. Surveys Tuts.}, 
  title={On the Road to {6G}: Visions, Requirements, Key Technologies, and Testbeds}, 
  year={2023},
  volume={25},
  number={2},
  pages={905-974},
  doi={10.1109/COMST.2023.3249835}}

@INPROCEEDINGS{mmwave_sensing1,
  author={Wang, Weizheng and Vaidya, Girish and Bhattacharjee, Anup and Fioranelli, Francesco and Zuniga, Marco},
  booktitle={Proc. of 19th DCOSS-IoT}, 
  title={A Long-Term Study Of mmWave Sensing In An Outdoor Urban Scenario}, 
  year={2023},
  volume={},
  number={},
  pages={240-247},
  doi={10.1109/DCOSS-IoT58021.2023.00049}}

@ARTICLE{mmwave_sensing3,
  author={Zhang, Jia and Xi, Rui and He, Yuan and Sun, Yimiao and Guo, Xiuzhen and Wang, Weiguo and Na, Xin and Liu, Yunhao and Shi, Zhenguo and Gu, Tao},
  journal={IEEE Commun. Surv. Tutorials.}, 
  title={A Survey of mmWave-Based Human Sensing: Technology, Platforms and Applications}, 
  year={2023},
  volume={25},
  number={4},
  pages={2052-2087},
  doi={10.1109/COMST.2023.3298300}}

@INPROCEEDINGS{THz_sensing1,
  author={Chaccour, Christina and Saad, Walid and Semiari, Omid and Bennis, Mehdi and Popovski, Petar},
  booktitle={Proc. of IEEE ICC}, 
  title={Joint Sensing and Communication for Situational Awareness in Wireless {THz} Systems}, 
  year={2022},
  volume={},
  number={},
  pages={3772-3777},
  doi={10.1109/ICC45855.2022.9838764}}

@ARTICLE{THz_sensing2,
  author={Jornet, Josep M. and Elayan, Hadeel and Nagatsuma, Tadao and Juntti, Markku and Pinto, Enrique T. R. and Kürner, Thomas and Guerboukha, Hichem and Mittleman, Daniel M. and Knightly, Edward},
  journal={IEEE Veh. Technol. Mag.}, 
  title={Mobile Terahertz Communication and Sensing Systems: A Future Look}, 
  year={2024},
  volume={19},
  number={4},
  pages={20-35},
  doi={10.1109/MVT.2024.3485258}}

@ARTICLE{11146803,
  author={Bai, Lu and Lu, Mengyuan and Huang, Ziwei and Cheng, Xiang},
  journal={IEEE Trans. Commun.}, 
  title={A Multi-Modal UAV-to-Ground Channel Model for 6G Intelligent Sensing-Communication Integration}, 
  year={2025},
  volume={},
  number={},
  pages={1-1},
  doi={10.1109/TCOMM.2025.3605450}}

@ARTICLE{yejian_mag,
  author={Lyu, Yejian and Yuan, Zhiqiang and Li, Mengting and Mbugua, Allan Wainaina and Kyösti, Pekka and Fan, Wei},
  journal={IEEE Communications Magazine}, 
  title={Enabling Long-Range Large-Scale Channel Sounding at Sub-THz Bands: Virtual Array and Radio-Over-Fiber Concepts}, 
  year={2024},
  volume={62},
  number={2},
  pages={16-22},
  doi={10.1109/MCOM.001.2200411}}

@inproceedings{strudel2021segmenter,
  title={Segmenter: Transformer for semantic segmentation},
  author={Strudel, Robin and Garcia, Ricardo and Laptev, Ivan and Schmid, Cordelia},
  booktitle={Proceedings of the IEEE/CVF international conference on computer vision},
  pages={7262--7272},
  year={2021}
}

@article{lyu2025hybrid,
  author={Y. Lyu and Z. Huang and S. Schwarz and C. Han},
  title={Hybrid Channel Modeling and Environment Reconstruction for Terahertz Monostatic Sensing},
  journal={IEEE Transactions on Wireless Communications},
  volume={24},
  number={10},
  pages={8492--8504},
  year={2025},
  doi={10.1109/TWC.2025.3567292}
}

@article{fang2025environment,
  author={Z.-T. Fang and Y.-J. Lyu and Z. Yu and C. Han},
  title={Environment Reconstruction in Terahertz Monostatic Sensing: Joint Millimeter-Level Geometry Mapping and Material Identification},
  journal={IEEE Journal of Selected Topics in Electromagnetics, Antennas and Propagation},
  volume={1},
  number={1},
  pages={265--277},
  year={2025},
  doi={10.1109/JSTEAP.2025.3605128}
}

@article{yang2023novel,
  author={R. Yang and C.-X. Wang and J. Huang and E.-H. M. Aggoune and Y. Hao},
  title={A Novel {6G} {ISAC} Channel Model Combining Forward and Backward Scattering},
  journal={IEEE Transactions on Wireless Communications},
  volume={22},
  number={11},
  pages={8050--8065},
  year={2023},
  doi={10.1109/TWC.2023.3258150}
}

@inproceedings{zhao2023bistatic,
  author={C. Luo and A. Tang and F. Gao and J. Liu and X. Wang},
  title={Channel Modeling Framework for Bistatic {ISAC} under {3GPP} Standard},
  booktitle={Proc. IEEE VTC-Spring},
  year={2024}
}

@inproceedings{yang2023localization,
  author={R. Yang and Y. Wu and J. Huang and C.-X. Wang},
  title={A Novel {3D} Non-Stationary Localization-Assisted {ISAC} Channel Model},
  booktitle={Proc. IEEE WCNC},
  pages={1--6},
  year={2023}
}

@article{zhang2023shared,
  author={Z. Zhang and R. He and B. Ai and M. Yang and X. Zhang and R. Chen and H. Zhang and Z. Zhong},
  title={A Shared Multipath Components Evolution Model for Integrated Sensing and Communication Channels},
  journal={IEEE Antennas and Wireless Propagation Letters},
  year={2023},
  note={early access}
}

@article{liu2024shared,
  author={Y. Liu and J. Zhang and Y. Zhang and Z. Yuan and G. Liu},
  title={A Shared Cluster-Based Stochastic Channel Model for Integrated Sensing and Communication Systems},
  journal={IEEE Transactions on Vehicular Technology},
  volume={73},
  number={5},
  pages={6032--6044},
  year={2024},
  doi={10.1109/TVT.2023.3337648}
}

@article{liu2024extend,
  author={Y. Liu and J. Zhang and Y. Zhang and H. Gong and T. Jiang and G. Liu},
  title={How to Extend {3D} {GBSM} to Integrated Sensing and Communication Channel with Sharing Feature?},
  journal={IEEE Wireless Communications Letters},
  volume={13},
  number={8},
  pages={2045--2049},
  year={2024}
}

@article{yang2024standardization,
  author={W. Yang and Y. Chen and N. Cardona and Y. Zhang and Z. Yu and M. Zhang and J. Li and Y. Chen and P. Zhu},
  title={Integrated Sensing and Communication Channel Modeling and Measurements: Requirements and Methodologies Toward {6G} Standardization},
  journal={IEEE Vehicular Technology Magazine},
  volume={19},
  number={2},
  pages={22--30},
  year={2024},
  doi={10.1109/MVT.2024.3383654}
}

@article{alkhateeb2023deepsense,
  author={A. Alkhateeb and G. Charan and T. Osman and A. Hredzak and J. Morais and U. Demirhan and N. Srinivas},
  title={{DeepSense} {6G}: A Large-Scale Real-World Multi-Modal Sensing and Communication Dataset},
  journal={IEEE Communications Magazine},
  volume={61},
  number={9},
  pages={122--128},
  year={2023},
  doi={10.1109/MCOM.006.2200730}
}

@inproceedings{charan2022vision,
  author={G. Charan and T. Osman and A. Hredzak and N. Thawdar and A. Alkhateeb},
  title={Vision-Position Multi-Modal Beam Prediction Using Real Millimeter Wave Datasets},
  booktitle={Proc. IEEE WCNC},
  pages={2727--2731},
  year={2022}
}

@article{mollah2025multimodality,
  author={M. B. Mollah and H. Wang and M. A. Karim and H. Fang},
  title={Multi-Modality Sensing in mmWave Beamforming for Connected Vehicles Using Deep Learning},
  journal={IEEE Transactions on Cognitive Communications and Networking},
  year={2025},
  doi={10.1109/TCCN.2025.3558026}
}

@article{mashhadi2021federated,
  author={M. B. Mashhadi and M. Jankowski and T. Y. Tung and S. Kobus and D. Gündüz},
  title={Federated mmWave Beam Selection Utilizing {LiDAR} Data},
  journal={IEEE Wireless Communications Letters},
  volume={10},
  number={10},
  pages={2269--2273},
  year={2021},
  doi={10.1109/LWC.2021.3099136}
}

@article{cheng2023intelligent,
  title={Intelligent multi-modal sensing-communication integration: Synesthesia of machines},
  author={Cheng, Xiang and Zhang, Haotian and Zhang, Jianan and Gao, Shijian and Li, Sijiang and Huang, Ziwei and Bai, Lu and Yang, Zonghui and Zheng, Xinhu and Yang, Liuqing},
  journal={IEEE Communications Surveys \& Tutorials},
  volume={26},
  number={1},
  pages={258--301},
  year={2023},
  publisher={IEEE}
}

@inproceedings{zhu2025raytracing,
  author={M.-J. Zhu and Y.-J. Lyu and C. Han},
  title={In-Vehicle Point-Cloud Based Ray-Tracing Channel Simulation and Wireless Planning in the Terahertz Band},
  booktitle={Proc. IEEE GLOBECOM},
  year={2025}
}

@article{cheng2025synthsom,
  title={SynthSoM: A synthetic intelligent multi-modal sensing-communication dataset for Synesthesia of Machines (SoM)},
  author={Cheng, Xiang and Huang, Ziwei and Yu, Yong and Bai, Lu and Sun, Mingran and Han, Zengrui and Zhang, Ruide and Li, Sijiang},
  journal={Scientific Data},
  volume={12},
  pages={819},
  year={2025}
}

@inproceedings{guan2020through,
  author = {J. Guan and W. Hu and W. Wu and Y. Zhang and J. Wang},
  title = {Through Fog High-Resolution Imaging Using Millimeter Wave Radar},
  booktitle = {Proc. IEEE/CVF Conf. on Computer Vision and Pattern Recognition (CVPR)},
  pages = {11464--11473},
  year = {2020},
  doi = {10.1109/CVPR42600.2020.01148}
}

@article{yu2024mobirfpose,
  author = {C. Yu and J. Wang and H. Liu and Y. Liu and Z. Zhang},
  title = {{MobiRFPose}: Portable RF-Based 3D Human Pose Camera},
  journal = {IEEE Transactions on Multimedia},
  volume = {26},
  number = {8},
  pages = {3715--3727},
  year = {2024},
  doi = {10.1109/TMM.2023.3314979}
}

@inproceedings{manjur2025multimodal,
    author    = {Manjur, Sultan Mohammad and Biswas, Sabyasachi and Gurbuz, Ali C.},
    title     = {A Multimodal Video and Radar Fusion Framework for High-Accuracy Isolated Sign Language Recognition},
    booktitle = {Proceedings of the IEEE/CVF International Conference on Computer Vision (ICCV) Workshops},
    month     = {October},
    year      = {2025},
    pages     = {5002-5011}
}

@Article{Guangzheng2023A,
  author   = {Jing, Guangzheng and Hong, Jingxiang and Rodríguez-Piñeiro, José and Yin, Xuefeng and Yu, Ziming and He, Jia and Li, Xianjin},
  journal  = {IEEE Wireless Commun. Lett.},
  title    = {Spatial Transformation Imaging in Terahertz Band: A Successive Interference Cancellation Approach},
  year     = {2023},
  number   = {8},
  pages    = {1414-1418},
  volume   = {12},
  doi      = {10.1109/LWC.2023.3276819},
  fjournal = {IEEE Wireless Communications Letters},
}

@Article{Guangzheng2025,
  author   = {Jing, Guangzheng and Hong, Jingxiang and Yin, Xuefeng and Rodríguez-Piñeiro, José and Tong, Yixiao and Yu, Yuning and Yu, Ziming},
  journal  = {IEEE Trans. Wireless Commun.},
  title    = {Measurement-based Channel Modeling for Spatial Non-stationarity with Multipath Components},
  year     = {2025},
  pages    = {1-1},
  doi      = {10.1109/TWC.2025.3612153},
  fjournal = {IEEE Transactions on Wireless Communications},
}

@ARTICLE{Fleury1999sage,
  author={Fleury, B.H. and Tschudin, M. and Heddergott, R. and Dahlhaus, D. and Ingeman Pedersen, K.},
  journal={IEEE Journal on Selected Areas in Communications}, 
  title={Channel parameter estimation in mobile radio environments using the {SAGE} algorithm}, 
  year={1999},
  volume={17},
  number={3},
  pages={434-450},
  doi={10.1109/49.753729}}

@ARTICLE{yejian_thz_mag,
  author={Han, Chong and Lyu, Yejian and Yu, Ziming and Wang, Guangjian and Wang, Cheng-Xiang},
  journal={IEEE Wireless Communications}, 
  title={Terahertz Integrated Sensing and Communication: Sensing Comes Before and For Communication}, 
  year={2026},
  volume={},
  number={},
  pages={1-7},
  doi={10.1109/MWC.2026.3673683}}

@ARTICLE{11364301,
  author={Ma, Shuangyuan and Zhang, Ruichen and Ma, Zhangfeng and Liu, Guangyuan and Niyato, Dusit and Ai, Bo and Zhang, Renmin},
  journal={IEEE Transactions on Vehicular Technology}, 
  title={Energy-Efficient Transmission in {STAR-RIS} Assisted Secure {ISAC} Networks With {RSMA}: An {MoE-RBPPO} Approach}, 
  year={2026},
  volume={},
  number={},
  pages={1-15},
  doi={10.1109/TVT.2026.3658351}}

\end{document}